\pdfoutput=1

\documentclass[journal]{IEEEtran}

\usepackage{color}
\usepackage{amsmath}
\usepackage{hyperref}

\ifCLASSINFOpdf
\usepackage[pdftex]{graphicx}
\DeclareGraphicsExtensions{.pdf,.jpeg,.png}
\else
\fi

\usepackage[tight,footnotesize]{subfigure}

\ifCLASSINFOpdf
\else
\fi

\begin{document}

\title{Neurostream: Scalable and Energy Efficient \\ Deep Learning with Smart Memory Cubes}

\author{Erfan~Azarkhish,
	Davide~Rossi,
	Igor~Loi,
	and~Luca~Benini,~\IEEEmembership{Fellow,~IEEE}%
	\thanks{E. Azarkhish, D. Rossi, and I. Loi are with the Department of Electrical, Electronic and Information Engineering, University of Bologna, 40136 Bologna, Italy (e-mails: \{erfan.azarkhish, davide.rossi, igor.loi\}@unibo.it).}%
	\thanks{L. Benini is with the Department of Information Technology and Electrical Engineering, Swiss Federal Institute of Technology Zurich, 8092 Zurich, Switzerland, and also with the Department of Electrical, Electronic and Information Engineering, University of Bologna, 40136 Bologna, Italy (e-mail: lbenini@iis.ee.ethz.ch).}%
	\thanks{This project has received funding from the European Union’s Horizon 2020 research and innovation programme (OPRECOMP) under grant agreement No 732631; Swiss National Science Foundation under grant 162524 (MicroLearn: Micropower Deep Learning), armasuisse Science \& Technology; and the ERC MultiTherman project (ERC-AdG-291125). The authors would like to thank Fabian Schuiki and Michael Schaffner for their help with training explorations.}%
}

\maketitle

\begin{abstract}
High-performance computing systems are moving towards 2.5D and 3D
memory hierarchies, based on High Bandwidth Memory (HBM) and Hybrid Memory Cube (HMC)
to mitigate the main memory bottlenecks. This trend is also creating new
opportunities to revisit near-memory computation. 
In this paper, we propose a flexible processor-in-memory (PIM) solution for scalable and energy-efficient execution of deep convolutional networks (ConvNets), one of the fastest-growing workloads for servers and high-end embedded systems. 
Our co-design approach consists of a network of Smart Memory Cubes (modular extensions to the standard HMC) each augmented with a many-core PIM platform called NeuroCluster. NeuroClusters have a modular design based on NeuroStream coprocessors (for Convolution-intensive computations) and general-purpose RISC-V cores.
In addition, a DRAM-friendly tiling mechanism and a scalable computation paradigm are presented to efficiently harness this computational capability with a very low programming effort.
NeuroCluster occupies only 8\% of the total logic-base (LoB) die area in a standard HMC and achieves an average performance of 240\,GFLOPS for complete execution of full-featured state-of-the-art (SoA) ConvNets within a power budget of 2.5\,W. Overall 11\,W is consumed in a single SMC device, with 22.5\,GFLOPS/W energy-efficiency
which is 3.5X better than the best GPU implementations in similar technologies. The minor increase in system-level power and the negligible area increase make our PIM system a cost-effective and energy efficient solution, easily scalable to 955\,GFLOPS with a small network of just four SMCs.

\end{abstract}


\IEEEpeerreviewmaketitle

\section{Introduction} \label{intro}

\begin{figure}[!t]
	\centering
	\begin{subfigure}
		\centering
		\includegraphics[width=0.90\columnwidth]{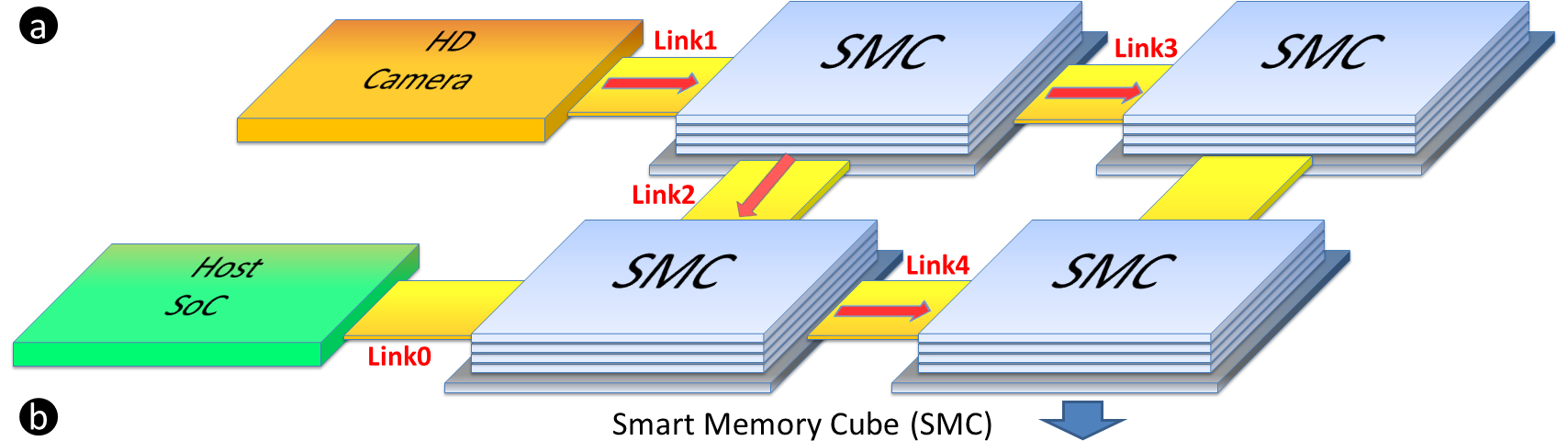}
	\end{subfigure}	
	\\
	\begin{subfigure}
		\centering
		\includegraphics[width=0.90\columnwidth]{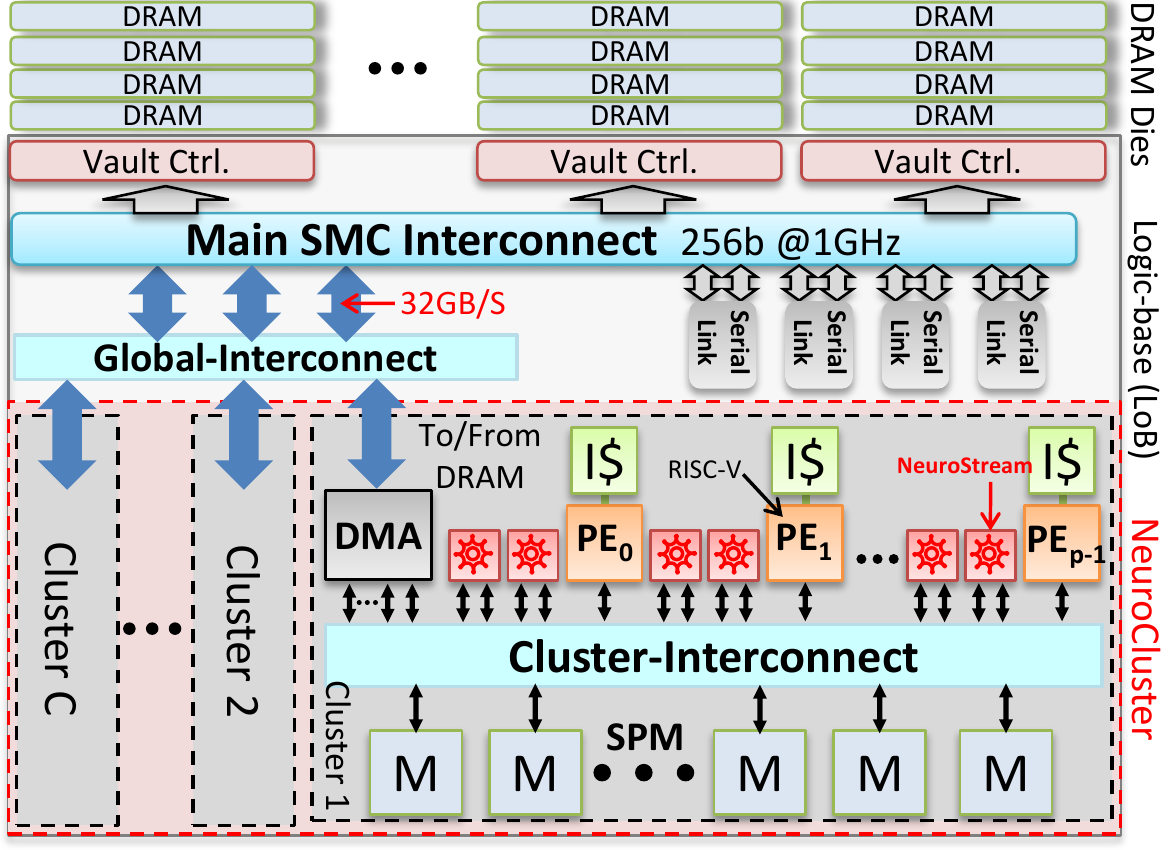}
	\end{subfigure}
	\\
	\vspace{-0.025\columnwidth}
	\begin{subfigure}
		\centering
		\includegraphics[width=0.93\columnwidth]{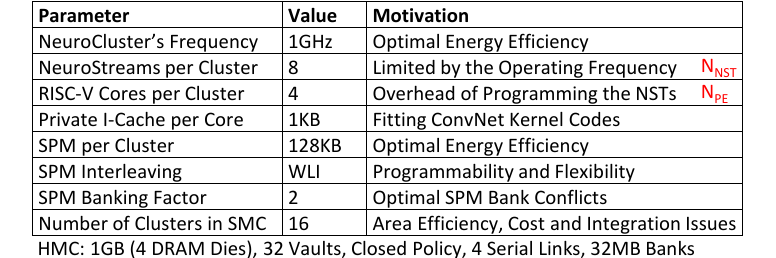}
	\end{subfigure}	
	\caption{(a) An overview of the SMC network for scalable ConvNet execution, (b) block diagram of one SMC instance highlighting the NeuroCluster platform along with the baseline system parameters.}
	\label{fig:sch}
\end{figure}

Today, brain-inspired computing (BIC) is successfully used in a wide variety of applications such as surveillance, robotics, industrial, medical, and entertainment systems. Recently, several research programs have been launched by major industrial players (e.g. Facebook, IBM, Google, Microsoft), pushing towards deploying services based on brain-inspired machine-learning (ML) to their customers \cite{GOOGLENET-PAPER}\cite{DEEPFACE}\cite{MICROSOFTCOCO}. These companies are interested in running such algorithms on powerful compute clusters in large data centers.
Convolutional neural networks (ConvNets) are known as the SoA ML algorithms specialized at BIC \cite{CONVNET-TAXONOMY}.
ConvNets process raw data directly, combining the classical models of feature extraction and classification into a single algorithm. The key advantages of them over traditional Multilayer-Perceptrons (MLP) are local connectivity and weight sharing:
Each neuron is connected only to a local region of the previous layer (or the input volume) called its receptive field \cite{VGGNETWORKS}. This is beneficial for dealing with high-dimensional inputs such as images. %
Moreover, weight sharing dramatically reduces the number of parameters that need to be stored.
ConvNets are not limited to image-processing only and they can be applied to other workloads such as audio and video \cite{CONVNET-VIDEO}, and even RFID-based activity recognition \cite{RFID}. %
Also, function approximation in scientific workloads \cite{Fathom} %
is another important target for ConvNets, motivating the need for a highly scalable and energy-efficient execution platform for them. In addition, recurrent networks (RNN) have been recently utilized for Deep Learning (DL) and implemented on scalable network-on-chips \cite{TPDS-SPIKING}\cite{TPDS-RNN}. These networks have a great potential for solving time-dependent pattern recognition problems because of their inherent dynamic representations. All these emerging DL models can be future targets for our PIM proposal, yet, in this paper, we focus on ConvNets for image and video.

A diverse range of ConvNet implementations exist today from standard software libraries running on general-purpose platforms \cite{CAFFE}\cite{CUDNN} to application-specific FPGA \cite{FPGA15}\cite{NEUFLOW}\cite{CAFFEINE}, ASIC \cite{EIE}\cite{DIANNAO}\cite{ORIGAMI}\cite{ShiDianNao}, and even initial explorations on near memory computing \cite{AMD}\cite{NEUROCUBE}\cite{PRIME}\cite{TETRIS}.
Even though ConvNets are computation-intensive workloads and extremely high energy-efficiencies have been previously reported for their ASIC implementations \cite{ORIGAMI}\cite{ShiDianNao}\cite{DIANNAO},
the scalability and energy-efficiency of modern ConvNets are ultimately bound by the main memory where their parameters and channels need to be stored (See \autoref{related-conv}). This makes them interesting candidates for near memory computation, offering them plenty of bandwidth at a lower cost and without much buffering compared to off-chip accelerators due to lower memory access latency (A consequence of the Little's law\footnote{Little's law ($L = \lambda W$) states that in a stable memory system, the long-term average buffer size ($L$) is equal to the long-term average effective bandwidth ($\lambda$) multiplied by the average memory access time ($W$).} \cite{ERFANARCS16}). 

Heterogeneous Three-dimensional (3D) integration is helping mitigate the well-known memory-wall problem \cite{ERFANTVLSI16} %
The Through-silicon-via (TSV) technology is reaching commercial maturity by memory manufacturers \cite{HMCSTANDARD}\cite{3DFLASH} to build memory cubes made of vertically stacked thinned memory dies in packages with smaller footprint and power compared with traditional multichip modules, achieving higher capacity. On the other hand, a new opportunity for revisiting near-memory computation to further close the gap between processors and memories has been provided in this new context \cite{AMD}\cite{NEUROCUBE}. %
This approach promises significant energy savings by avoiding energy waste in the path from processors to memories.
In 2013, an industrial consortium backed by several major semiconductor companies standardized the hybrid memory cube (HMC) \cite{HMCSTANDARD} as a modular and abstracted 3D memory stack of multiple DRAM dies placed over a logic base (LoB) die, providing a high-speed serial interface to the external world.
More recently, a fully backward compatible extension to the standard HMC called the smart memory cube (SMC) was introduced in \cite{ERFANTVLSI16} along with a flexible programming-model \cite{ERFANARCS16}, augmenting the LoB die with generic PIM capabilities. %

In this paper, we propose a scalable, flexible, and energy-efficient platform targeting large-scale execution of deep ConvNets with growing memory footprints and computation requirements. Our proposal increases the total LoB die area of a standard HMC only by 8\% and achieves 240\,GFLOPS on average for complete execution of full-featured ConvNets within a power-budget of 2.5\,W. 22.5\,GFLOPS/W energy efficiency is achieved in the whole 3D stack (consuming 11\,W in total) which is 3.5X better than the best GPU implementations in similar technologies. %
We also demonstrate that using a flexible tiling mechanism along with a scalable computation paradigm it is possible to efficiently utilize this platform beyond 90\% of its roofline \cite{ROOFLINE} limit, and scale its performance to 955\,GFLOPS with a network of four SMCs. We have adopted the cycle-accurate SMC model previously developed in \cite{ERFANTVLSI16} along with the generic software stack provided in \cite{ERFANARCS16}, and built a Register-Transfer-Level (RTL) model for our DL framework, along with the required software layers.
Our main contributions can be summarized as follows: I) Using near memory computation for large-scale acceleration of deep ConvNets with large memory footprints, requiring the use of DRAM; II)	Proposing the NeuroStream coprocessors as an alternative to vector-processing, providing a flexible form of parallel execution without the need for fine-grained synchronization; III) Presenting a flexible tiling mechanism and a scalable computation paradigm for ConvNets, achieving more than 90\% roofline utilization; IV) A low-cost and energy-efficient implementation of this solution based on a standard HMC device, scalable to a network of multiple HMCs.

This paper is organized as follows. Background and related work are presented in \autoref{related}. Our architectural design methodology, computation paradigm, and programming model are explained in Sections \autoref{arch}, \autoref{comp-model}, and \autoref{prog-model} respectively. Experimental results are in \autoref{exp}. Conclusions and future directions are explained in \autoref{con}.

\section{Background and Related Work} \label{related}

A brief introduction to ConvNets is presented in \autoref{background}. The evolution of modern ConvNets and their uprising implementation challenges are explained in \autoref{related-conv}. The existing implementations for them are compared with this work in \autoref{related-impl}.

\subsection{Convolutional Neural Networks} \label{background}

ConvNets are typically built by repeated concatenation of five classes of layers: convolutional (CONV), activation (ACT), pooling (POOL), fully-connected (FC), and classification (CLASS) \cite{DLBOOK}. 
CONV is the core building block of the ConvNets doing most of the computational heavy-lifting for feature extraction. It essentially consists of Multiply-and-accumulate (MAC) operations as shown below \cite{DLBOOK}:

\begin{small}
	\begin{align}
	&y_{o}^{l}(i,j) = b_o^{l} + \sum_{c \in C_{i}} \sum_{(a,b) \in K} k_{o,c}^l(b,a) x_{c}^l(j-b,i-a)\nonumber
	\end{align}
\end{small}

where $o$ indexes the output channels ($C_{o}^l$), $c$ indexes the input channels ($C_{i}^l$), and $K$ denotes the convolution kernels (a.k.a filters).
After each CONV layer, a non-linear activation function (e.g. \textit{sigmoid}, \textit{tanh}, or \textit{ReLU} \cite{DLBOOK}) is applied to the output $y$ of each individual neuron. This non-linearity gives neural networks (NNs) superior classification and learning capabilities over linear classifiers and allows them to solve non-trivial problems.
\textit{Sigmoid} and \textit{tanh} come from the traditional multilayer perceptrons, and their requirement for the computation of exponential functions makes them unsuitable for the main activation function \cite{DLBOOK}. In modern feed-forward NNs the common recommendation is to use the rectified linear unit (\textit{ReLU}) defined by $g(z)=max\{0,z\}$. Applying this function to the output of a linear transformation yields a piecewise-linear function. For this reason, it preserves many of the properties that make linear models easy to generalize and optimize with gradient-based methods \cite{DLBOOK}.
It is common to periodically insert a POOL layer in-between successive CONV layers. Its function is to progressively reduce the size of the volume (e.g. by calculating the maximum value of every $k{\times}k$ region). This is to reduce the amount of parameters and computation in the network and to control over-fitting \cite{DLBOOK}.
In the final layers, multiple FC layers and one CLASS layer perform the final classification and transform the results into several classes. FC layers have a full connectivity and work similar to MLPs.
The CLASS layer converts the outputs of the network to categorical distributions. A widely used classifier is the SoftMax function. Compared to the rest of the network, its computational complexity is usually small \cite{CONVNET-TAXONOMY}\cite{LUKAS-DAC15}.
The first layer connects the network to the input volume which can be an image, a video frame, or a signal, depending on the application (a 3-channel R,G,B image for instance). Each layer $l$ transforms the input volume $(X_i, Y_i, C_i)$ into an output volume $(X_o, Y_o, C_o)$. This terminology is used throughout this paper and will be further elaborated in \autoref{4dtile}.

\subsection{Implementation Challenges of Modern ConvNets} \label{related-conv}

\begin{table}[!t]
	\centering
	\caption{Storage requirement (MB) in the SoA ConvNets.}
	\includegraphics[width=0.85\columnwidth]{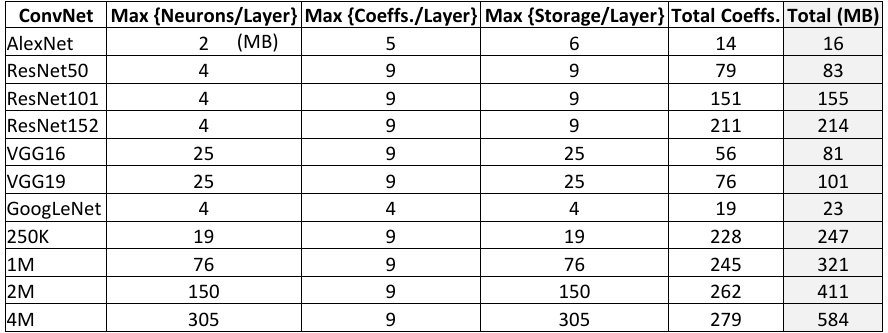}
	\label{tbl:storage}
\end{table}

ConvNets have been rapidly evolving in the past years, from small networks of only a few layers (e.g. LeNet-5 \cite{DLBOOK}) to over hundred \cite{RESNET-PAPER} and thousand \cite{RESNET1K} layers, and from having a few kilobytes of coefficients (a.k.a. weights) to multi-mega bytes in \cite{GOOGLENET-PAPER}\cite{VGGNETWORKS}\cite{RESNET-PAPER}.
Also, traditional ConvNets were only applicable to small 32x32 images, while the SoA ConvNets have 224x224 inputs, and this size is expected to grow \cite{DLBOOK}.
\autoref{tbl:storage} shows an estimation for the storage requirements (in MB) of top-performing ConvNets, assuming layer-by-layer execution. AlexNet \cite{DLBOOK} is the 2012 winner of the ILSVRC challenge \cite{LSVRC}. VGG networks \cite{VGGNETWORKS} and GoogLeNet \cite{GOOGLENET-PAPER} were the winners of different categories in 2014, and ResNet \cite{RESNET-PAPER} was the most recent winner of this challenge in 2015. ResNet1K with 1001 layers \cite{RESNET1K} is omitted from our study because its training loss and validation error (for the ImageNet database \cite{LSVRC}) are not yet lower than its previous versions. Instead in this paper, ResNet-152 is extended to larger networks (accepting {250K}/{1M}/{2M}/{4M}-pixel images shown in  \autoref{tbl:storage}) to further investigate the scalability of our approach and its applicability to beyond High-Definition (HD) image resolutions. ResNet is chosen for this purpose because it is more challenging to accelerate than the other networks (See \autoref{perf}).

It can be clearly seen that the typical on-chip (L1, L2) storages in the memory hierarchy (caches or SRAM-based scratchpad memories) cannot accommodate even a single layer of these ConvNets, as the required storages per layer range from 6\,MB to over 300\,MB. In addition, the assumption that all coefficients can be stored on-chip (\cite{EIE}\cite{DIANNAO}\cite{NEUFLOW-ASIC}) is not valid anymore, since an additional storage of 14\,$\sim$\,280\,MB is required to accommodate the coefficients. Overall, 16$\sim$580\,MB is needed for layer-by-layer execution, demonstrating that DRAM is necessary as the main storage for deep ConvNets and also motivating computation near main memory. %
A similar observation was recently made in \cite{NEUROCUBE}.

Another point is that the straightforward topology of the traditional ConvNets such as LeNet-5 has recently evolved to more complex topologies such as Deep Residual Learning in ResNet \cite{RESNET-PAPER} and the Inception Model (Network in Network) in GoogLeNet \cite{GOOGLENET-PAPER}. This makes application specific implementations less practical and highlights the need for flexible and programmable platforms.
Also, unlike traditional ConvNets with very large and efficient convolution filters (a.k.a. feature maps) of over 10x10 inputs, modern ConvNets tend to have very small filters (e.g. 3x3 in VGG and 1x1 in GoogLeNet and ResNet). It can be easily verified that the Operational Intensity (OI)\footnote{Operational Intensity (OI), a.k.a. Computation to Communication Ratio, is a measure of computational efficiency defined in the roofline-model \cite{ROOFLINE} as the number of computations divided by the total transferred data (bytes).} \label{DEFINITON-OI}
decreases as the convolution filters shrink. This can negatively impact computation, energy, and bandwidth efficiency (See \autoref{exp}). In this paper, we design a scalable PIM platform capable of running very deep networks with large input volumes and arbitrary filter sizes.

Lastly, different tiling methods for ConvNets have been previously proposed \cite{FPGA15}\cite{CAFFEINE} %
for FPGA implementations, in \cite{DIANNAO} for a neuromorphic accelerator, and in \cite{NVE} for a Very Long Instruction Word (VLIW) architecture. In \cite{NVE} a tile-strip mechanism is proposed to improve locality and inter-tile data reuse for ConvNets with large filters. In \cite{CAFFEINE} a row-major data layout has been proposed to improve DRAM's bandwidth efficiency and reduce bank conflicts in FPGA's BRAM banks.
Also, tile-aware memory layouts have been previously proven effective for multi-core \cite{SVD} and GPU implementations \cite{TPDS-GPUTRANSPOSE} of linear algebra algorithms, directly affecting their cache performance, bandwidth efficiency, and the degree of parallelism.
In this paper, we introduce a general and flexible form called 4D-tiling (\autoref{4dtile}) allowing for optimization of performance and energy efficiency under given constraints such as on-die SPM and DRAM bandwidth usage. Our proposed mechanism reduces the communication overheads among the clusters and uses the DRAM interface more efficiently by merging DMA transfers into larger chunks.

Throughout this paper, we use single-precision floating-point (FP32) arithmetic to be able to flexibly target large-scale DL in the high-performance computing domain. The wide dynamic range offered by this representation improves programmability and allows for implementing a wider range of algorithms, as well as, training and backpropagation, since they usually require higher precision and dynamic range \cite{DLBOOK}.
We use the notion of GFLOPS (Giga-FLOPS per second) to demonstrate the achieved FP32 performance, along with GOPS (Giga-operations per second) to show integer/fixed-point performance in \autoref{related-impl}.

\subsection{SoA ConvNet Implementations}\label{related-impl}

A glance at the SoA highlights two main directions:
(I) Application-specific architectures based on ASIC/FPGAs \cite{NEUFLOW-ASIC}\cite{ORIGAMI}\cite{DIANNAO}\cite{FPGA15}\cite{CAFFEINE}\cite{EIE}\cite{EYERISS};
(II) Software implementations on programmable general-purpose platforms such as CPUs and GPUs \cite{LUKAS-DAC15}\cite{CAFFE-GPU}\cite{FPGA15}\cite{CAMBRICON}.
ASIC ConvNet implementations achieve impressive energy efficiency and performance:
DianNao \cite{DIANNAO} obtains 450\,GOPS at 0.5\,W with a neuromorphic architecture using 16b fixed-point arithmetic in 65nm technology. Later, it has been extended to 1250\,GOPS within a similar power budget in \cite{ShiDianNao}.
The limiting assumption in this work is that the whole ConvNet (coefficients + the largest intermediate layer of LeNet-5) fits inside the on-chip SRAM ($\sim$256kB). As we showed above, this assumption is not valid anymore for modern ConvNets. Also, they use a small input image size (32x32) with very large convolution filters (e.g. 18x18, 7x7), which is unrealistic for modern ConvNets, as explained before.
In EIE \cite{EIE} coefficients are compressed by pruning and weight-sharing, achieving 100\,GOPS at 625\,mW in 45nm technology, with the main drawback of storing 84M coefficients on-chip, resulting in an area of over 40$mm^{2}$.
Eyeriss \cite{EYERISS} presents a reconfigurable ConvNet accelerator mainly focusing on reducing data movement by an approach called ``row-stationary'' computation, in which kernel coefficients are loaded once and reused several times. Eyeriss achieves around 70\,GOPS at 278\,mW for AlexNet, but when scaling to VGG16, their performance drops to 20\,GOPS within the same power budget. In \cite{TETRIS} it is shown that memory is the main bottleneck of Eyeriss, limiting its scalability and energy efficiency when used with larger networks and images.
Origami \cite{ORIGAMI} achieves 145\,GOPS at 0.5\,W, using 12b fixed-point implementation (65nm-UMC technology at 1.2V, with 40kB of storage), being scalable to 800\,GOPS/W at 0.8V.
The main issue with these works is their lack of flexibility and scalability to large inputs and modern ConvNets. Also, the assumption that a significant part of the ConvNet can be stored on-chip is not valid anymore, and shrinking filter dimensions 
can significantly hurt their reported performance and efficiency numbers with 18x18 filters in \cite{DIANNAO}, 10x10 in \cite{NEUFLOW-ASIC}, 7x7 in \cite{NEUROCUBE}, and 6x6 in \cite{ORIGAMI}, due to the significantly reduced OI.
In this paper, we propose a flexible solution supporting a wide range of ConvNets with different network, kernel, and image dimensions.
FPGA platforms provide higher flexibility compared to ASIC implementations but lower energy/area efficiency due to the usage of reconfigurable routing switches and logic blocks.
In \cite{FPGA15}, ConvNet models are synthesized to Xilinx Virtex7-485T using high-level synthesis. 61\,GFLOPS is achieved (FP32) at 18\,W (3.4\,GFLOPS/W). In \cite{NEUFLOW-ASIC} the NeuFlow data-flow vision processor has been prototyped on Xilinx Virtex-6 VLX240T and 147\,GOPS @ 10\,W (14.7\,GOPS/W) is achieved. 
Caffeine \cite{CAFFEINE} presents a flexible hardware/software co-design library to efficiently accelerate ConvNets on FPGAs. It achieves 166\,GOPS @ 25\,W (6.6 GOPS/W) on Xilinx KU060 and 8.5 GOPSW on Xilinx VX690T with 16b fixed-point arithmetic. In comparison with CPU/GPU platforms, low-cost FPGAs have limited memory bandwidth which is also highly sensitive to memory access burst lengths, requiring a more careful design for efficient bandwidth usage. High-end FPGAs offer larger bandwidths thanks to their larger number of high-speed IOs. The problem is that these IOs are very general (because of the reconfigurability requirements) and therefore they are very expensive in area and power \cite{CAFFEINE}.
Our proposal achieves higher energy-efficiency thanks to near memory computation and having optimized DMA interfaces to DRAM with a novel tiling scheme.
In addition, the higher bandwidth available to our solution translates into lower programming effort (according to the roofline model \cite{ROOFLINE}) and reasonable performance, even for applications not super-optimized to use the available bandwidth efficiently.
General-purpose GPU platforms, on the other hand, are able to flexibly execute different deep NNs \cite{CAFFE-GPU}\cite{LUKAS-DAC15}\cite{FPGA15} without the limitations of application specific architectures.
Fast and user-friendly frameworks such as CAFFE \cite{CAFFE} and cuDNN \cite{CUDNN} are publicly available which also provide facilities to efficiently train deep NNs.
In \cite{CAFFE-GPU} over 500\,GFLOPS has been reported for execution of the CAFFE models based on cuDNN on NVIDIA Tesla K40 with default settings. By turning off error-correction and boosting the clock speed they have been able to reach 1092\,GFLOPS @235\,W (4.6\,GFLOPS/W). Geforce GTX 770 achieves 2.6\,GFLOPS/W using the same framework \cite{CAFFE-GPU}.
Mobile GPUs achieve similar energy efficiencies at lower power budgets. 54\,GFLOPS for less than 30\,W is reported in \cite{NEUFLOW-ASIC} for  NVIDIA GT335M, and in \cite{LUKAS-DAC15} 84\,GFLOPS for 11\,W is reported for NVIDIA Tegra K1.
More recently NVIDIA \cite{NVIDIA-WP17} has reported promising energy and performance improvement for its high-end GPU accelerator Tesla P100 in 16nm technology and with a new framework called TensorRT which is 1.5X more efficient than CAFFE. For inference with GoogLeNet, ResNet-50, and AlexNet, 20, 23.9, and 35\,GFLOPS/W are reported, respectively.
We would like to remind here that Tesla P100 is an expensive high-end accelerator costing more than \$9K, while our PIM solution can be integrated within existing systems with HMC devices at almost no additional cost, in the same package structure, and within the same power budget. Plus, an HMC module itself costs less than \$1.4K, which is expected to reduce as its market size grows. %
CPU implementations achieve lower energy efficiency for execution of ConvNets with standard frameworks. In \cite{FPGA15}, 12.8\,GFLOPS at 95\,W has been reported for Intel Xeon CPU E5-2430 (@2.20GHz) with 15MB cache and 16 threads. In \cite{LUKAS-DAC15}, 35\,GFLOPS at 230\,W has been reported for Intel Xeon E5-1620v2.
In \cite{CAMBRICON} a domain-specific instruction set architecture (ISA) is designed for the widely used NN models by identifying the common operations in them. They show higher flexibility compared to \cite{DIANNAO} by being able to model 9 classes of NNs. The size of the studied networks, however, is extremely small compared to the ones studied in our paper.
Another common approach is to augment a RISC processor with Single-Instruction-on-Multiple-Data (SIMD) extensions. %
Commercial platforms such as TI AccelerationPAC, CEVA-XM4, Synopsys DesignWare EV5x, and Movidius Fathom %
follow this trend.
Performance and efficiency characterization of these platforms is not publicly available, nevertheless, SIMD extensions require more programming effort to be efficiently utilized, and their register-file bottleneck limits their scalability \cite{CAMBRICON}. In this paper, we follow a different approach based on many scalar coprocessors working in parallel on a shared memory. This is described in \autoref{arch}.
On the other hand, Google's TensorFlow platform \cite{TENSORFLOW} maps large-scale ML problems to several machines and computation devices, including multi-core CPUs, general-purpose GPUs, and custom designed ASICs known as Tensor Processing Units (TPUs).
Nervana, also, has built a scalable ML platform \cite{NERVANA} with their own implementation of TPUs, and a library called Neon to support cloud computation with different back-ends. Apache Spark features a library called MLlib \cite{MLLIB} targeting scalable practical ML. No performance or efficiency data is publicly available for these platforms. Lastly, HCL2 \cite{TPDS-HANDOOP} motivates designing a heterogeneous programming system based on map-reduce for ML applications supporting CAFFE \cite{CAFFE} representations.

The study of the ConvNets in a near-memory context has been done in \cite{AMD}\cite{NEUROCUBE}\cite{PRIME}\cite{TETRIS}.
In \cite{AMD} the authors assume that the whole internal bandwidth of the HMC (320\,GB/s) is available to PIM. They reach a performance of 160\,GFLOPS (lower compared to our solution) for AlexNet and VGG inside each cube,
and the details of their PIM design are not exposed in their work.  Plus, instead of performance efficiency, normalized execution time is reported only, and the analysis of power and area are left as future works.
In \cite{NEUROCUBE} a data-driven computing model is proposed using finite-state-machines (FSM) near each HMC vault controller, preprogrammed to generate DRAM addresses for the ConvNet under execution (16b fixed-point). Their study, however, is limited to a small ConvNet with 6 layers and scaling their approach to modern ConvNets seems difficult. They achieve 132\,GOPS @ 13\,W with an energy efficiency lower compared to our work (10\,GOPS/W). The LoB die in NeuroCube consumes 3.4\,W, mainly due to the presence of data caches, on-chip storage for weights, and network-on-chip routers with packet encapsulation in their accelerator design.

Tetris \cite{TETRIS} is a scalable NN accelerator based on HMC. It uses the ``row-stationary'' computation paradigm proposed in \cite{EYERISS} with fixed-point computation and scales it to multiple NN engines each associated with a DRAM vault. Tetris requires an area of 3.5$mm^2$ per vault in the 45nm technology, which can be scaled to 21$mm^2$ in 28nm technology. From the relative results reported in \cite{TETRIS} its performance can be estimated as 159\,GOPS with an average power consumption of 6.9\,W. Both energy and area efficiency of Tetris are lower than our work.

Finally, in \cite{PRIME}, ConvNet execution in Re-RAM based non-volatile memory is investigated with different design decisions due to the drastically different memory technology used. Relative performance and energy numbers reported in this work make it difficult to compare directly, nevertheless, a throughout survey on the techniques to use these memories in comparison with DRAM is presented in \cite{TPDS-NVSURVEY}.
In this paper, we have observed that for modern ConvNets with shrinking kernels, coefficient reuse is becoming less practical and approaches such as row-stationary are not that beneficial anymore. For this reason, we use a completely different approach focusing on parallelism rather than coefficient reuse.

To summarize, three main assumptions motivate our proposed computation paradigm and tiling mechanism: a) Focusing on synchronization-free parallelism rather than coefficient reuse; b) Limiting the on-chip storage available to the PIM cluster; c) Supporting very large input images (up to 32Mega-pixels). We will demonstrate that our scalable and flexible ConvNet acceleration platform provides higher energy efficiency compared to the best FPGA and GPU implementations in similar technologies at a fraction of their system cost.

\section{System Architecture} \label{arch}

ConvNets, by nature, are computation demanding algorithms. One forward pass of VGG19, for example, requires around 20 billion MAC operations with over 100K operations per pixel. Maintaining even a frame-rate of 10 frames per second will require over 200 GFLOPS.
In theory, ConvNets can reach extremely high OI ratios (discussed in \autoref{related-conv}), as they reuse data efficiently. However, due to the very large memory footprints of deep ConvNets, their performance and energy efficiency is ultimately constrained by the main DRAM storage and off-chip communication. As we will show throughout this paper, in a near-memory context some of these constraints can be relaxed, providing the possibility to improve energy efficiency and programmability.
\autoref{neurocluster} describes the design of our many-core PIM platform.

\subsection{NeuroCluster} \label{neurocluster}

NeuroCluster (Illustrated in \autoref{fig:sch}b) is a flexible general purpose clustered many-core platform, designed based on energy-efficient RISC-V processing-elements (PEs) \cite{RISCV} and NeuroStream (NST) coprocessors (described in \autoref{neurostream}), all grouped in tightly-coupled clusters. Each cluster consists of four PEs and eight NSTs, with each PE being responsible for programming and coordinating two of the NSTs. This configuration is found to be optimal in the explorations presented in \autoref{exp}.
The PEs are augmented with a light-weight Memory Management Unit (MMU) along with a small sized Translation Look-aside Buffer (TLB) providing zero-copy virtual pointer sharing from the host to NeuroCluster (More information in \autoref{prog-model}).
Instead of caches and prefetchers which provide a higher level of abstraction without much control, and they are more suitable for host-side accelerators \cite{ERFANARCS16}, scratchpad memories (SPMs) and DMA engines are used with a simple and efficient computation paradigm to boost energy efficiency \cite{Rossi2016}\cite{ERFANARCS16}\cite{NVE}. Also, caches introduce several coherence and consistency concerns and are less area and energy-efficient in comparison with SPMs \cite{ERFANARCS16}. %
Each cluster features a DMA engine capable of performing bulk data transfers between the DRAM vaults and the SPM inside that cluster. It supports up to 32 outstanding transactions and accepts virtual address ranges without any alignment or size restrictions. The NST coprocessors, on the other hand, have limited visibility only to the cluster's SPM with no concerns about address translations and DMA transfers.
This mechanism allows for simple and efficient computation while maintaining the benefits of virtual memory support \cite{ERFANARCS16}.

Each PE is a light-weight RISC-V based processor with 4 pipeline stages and in-order execution (without branch prediction, predication, or multiple issue) for energy-efficient operation \cite{RISCV}. RTL models of these cores have been adopted from \cite{PULP}.
1\,kB of private instruction-cache (4-way set associative) is available to each core. %
An in-depth exploration of different instruction cache choices (including size, associativity, and shared/private organizations) are previously performed in \cite{IGOR-ICACHE}, demonstrating that this organization not only supports larger data-sets (e.g. ConvNets),
but also larger codes, as long as their main computing loops (kernels) fit in the caches.
The SPM inside each cluster is word-level-interleaved (WLI) with multiple banks accessible through the cluster-interconnect.
The cluster-interconnect has been designed based on the logarithmic-interconnect proposed in \cite{ERFANTVLSI14} to provide low-latency all-to-all connectivity inside the clusters. Also, the AXI-4 based global-interconnect, connecting the clusters, follows the same architecture as the SMC-Interconnect \cite{ERFANTVLSI16} to achieve a very high bandwidth.

\subsection{NeuroStream} \label{neurostream}

NeuroStream (NST) is a streaming coprocessor designed based on two observations: (I) Modern ConvNets tend to have very small convolution filters, making coefficient reuse less practical (previously discussed in \autoref{related-conv}). (II) The most demanding operation in ConvNets is MAC \cite{LUKAS-DAC15}.
Therefore, unlike conventional SIMD coprocessors (e.g. ARM NEON), NST works directly on the shared multi-bank SPM without having many internal registers (just one accumulator). This feature along with its dedicated hardware address generators allows it to perform arbitrary computations efficiently and directly on the SPM. This removes the register-file bottleneck which is present in SIMD architectures and allows it to achieve a performance close to 1 MAC/cycle. Moreover, each NST can be treated as a scalar coprocessor working independently. Yet, it is possible to instantiate several NSTs inside a cluster to achieve a scalable parallelism without the need for fine-grained synchronization among them. This way, NSTs are easier to program compared to SIMD units, and they offer more flexibility in terms of the size/shape/stride of the computations.
In total, 128 instances of NST, clocked at a moderate speed of 1\,GHz, sum up to 256\,GFLOPS of raw performance in the NeuroCluster.

\begin{figure}[!t]
	\centering
	\includegraphics[width=0.9\columnwidth]{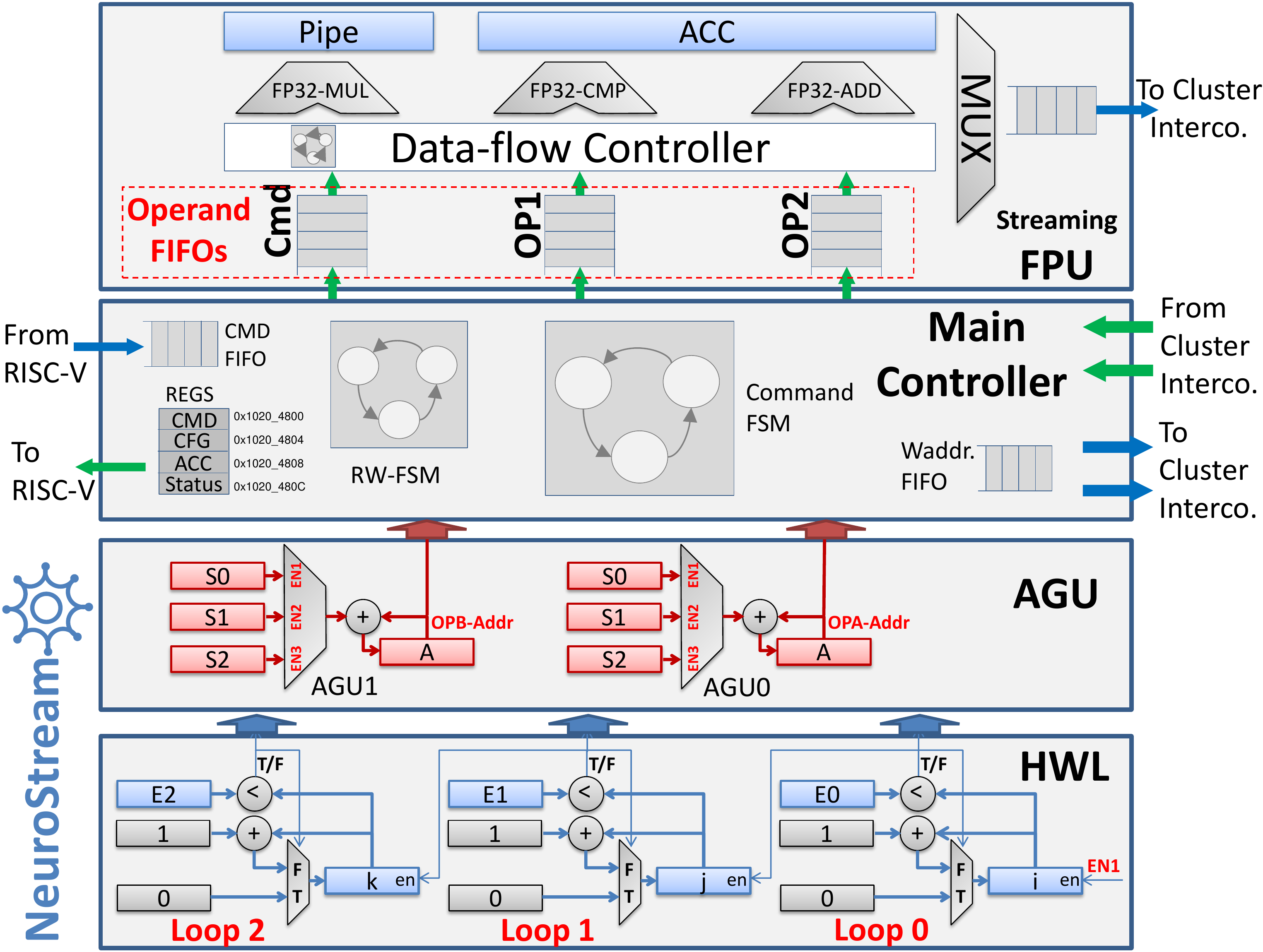}
	\caption{Architecture of the NeuroStream (NST) floating-point coprocessors.}
	\label{fig:nst}
\end{figure}

\autoref{fig:nst} illustrates the block diagram of NST, composed of the main controller, three hardware-loops (HWL), two Address Generation Units (AGUs), and an FP32 datapath (FPU) compatible with the IEEE-754 standard.
The main-controller is responsible for receiving the commands from the processor and issuing them to the datapath. A parametric-depth first-in-first-out (FIFO) command-queue is implemented to hide the programming latencies. Also, the control interface is memory-mapped, making it possible for the NSTs to easily communicate with other processor micro-architectures (e.g. ARM).
NSTs follow a nonblocking data-flow computation paradigm, and information flows in them as tokens. The main controller is, therefore, responsible for issuing enough transactions (2 in each cycle in case of MAC) towards the SPM and filling up the operand FIFOs to keep the FPU busy almost every cycle.
The hardware-loops are programmable FSMs capable of modeling up to three nested-loops in hardware. The AGUs can be programmed to generate complex strided SPM access patterns (See \autoref{prog-nst}).
By having two direct ports to the cluster-interconnect, each NST can fetch two operands (typically one coefficient and one data) in a single-cycle and perform an operation on them.

NST supports strided convolution, max-pooling, ReLU-activation, along with some basic utilities for backpropagation and training. Apart from these tasks, it can also be used for generic computations such as dot product, matrix multiplication, linear transformations, and weighted sum/average. Even single FP32 operations (e.g. add, multiply) are supported for generality. More than 14 commands in three categories are implemented: streaming (e.g. \textit{STREAM\_MAC}, \textit{STREAM\_SUM}, \textit{STREAM\_MAX}), single (e.g. \textit{SINGLE\_ADD}, \textit{SINGLE\_MUL}), and memory commands (for configuration and memory transfers to/from the accumulator). \autoref{prog-nst} describes how NSTs can be programmed to do various computations.

\section{Computation Model} \label{comp-model}

When a ConvNet such as GoogLeNet is selected for execution over our PIM system, first it is tiled using the 4D-tiling mechanism described in \autoref{4dtile}. This procedure prepares it for parallel execution over the clusters, and optimally partitions it to achieve the highest efficiency under given constraints such as on-die SPM and DRAM bandwidth usage.
Next, all coefficients are loaded in SMC's DRAM and an additional space is reserved there for the intermediate results of the largest layer (shown previously in \autoref{tbl:storage}). The input volume (e.g. the image or video frame) is loaded into this area before each run. The actual execution takes place layer-by-layer, each layer being parallelized over 16 clusters. Each cluster executes one 4D-tile at a time with all its NSTs working cooperatively to compute its final result inside the cluster's SPM. Only at the end of each layer, the clusters are synchronized. %
A more detailed description follows in \autoref{4dtile} and \autoref{mapping}.

\subsection{4D-Tiling Mechanism} \label{4dtile}

\begin{figure}[!t] %
	\centering
	\includegraphics[width=0.85\columnwidth]{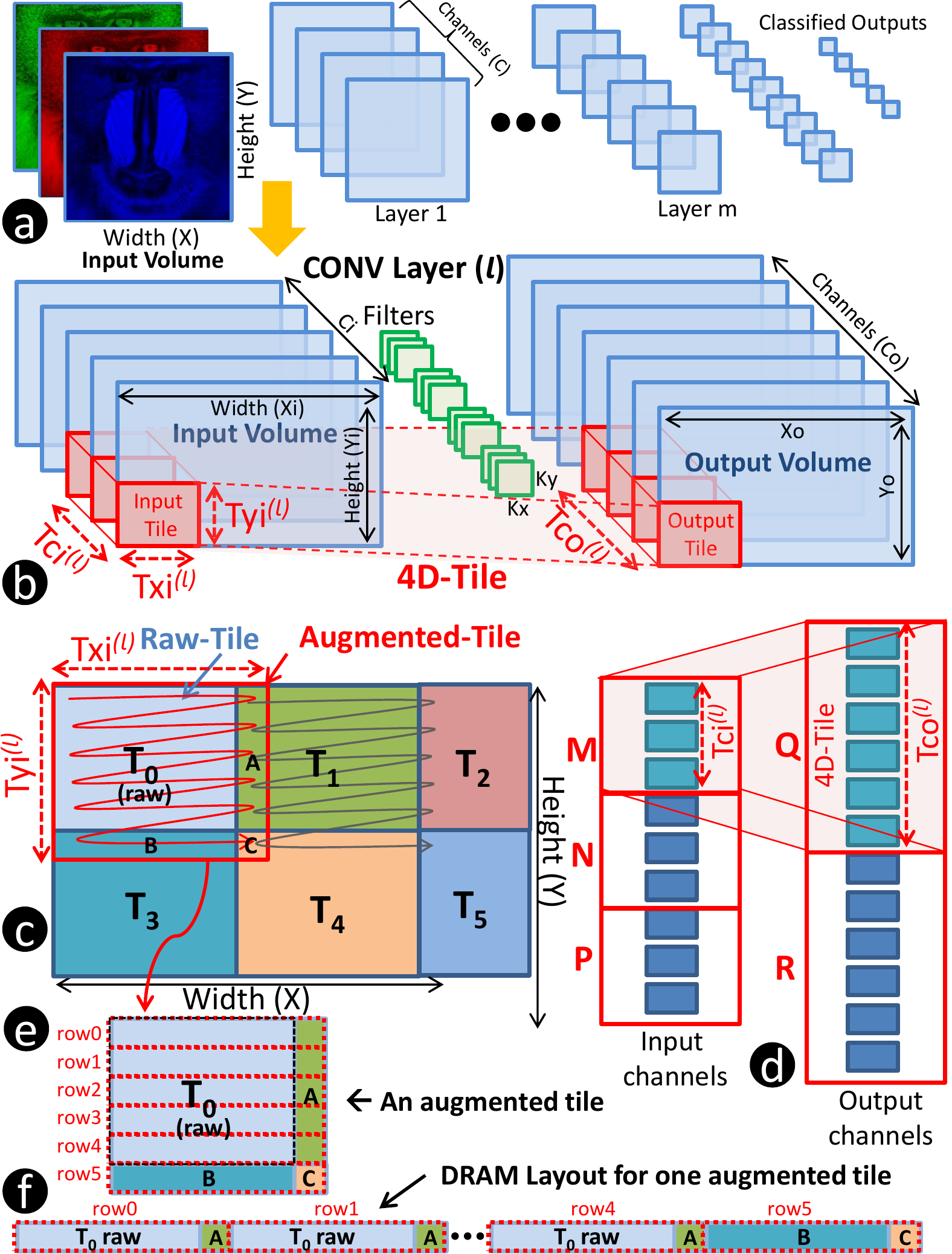}
	\caption{(a) Illustration of a general ConvNet, (b) a 4D-tile, (c) row-major data layout and tile-overlapping, (d) partial computation of tiles, and (e,f) the underlying DRAM storage of one augmented-tile.}
	\label{fig:tile4d}
\end{figure}

A 4D-tile (illustrated in \autoref{fig:tile4d}a,b) is a subset of the input volume (called Input-tile) and output volume (Output-tile) of a convolutional layer (\textit{l}) identified by the ($T_{Xi}^{(l)}$, $T_{Yi}^{(l)}$, $T_{Ci}^{(l)}$, $T_{Co}^{(l)}$) tuple. $T_{Xi}^{(l)}$ and $T_{Yi}^{(l)}$ are the tile width and height of the input volume of layer ${l}$, and $T_{Ci}^{(l)}$ and $T_{Co}^{(l)}$ are the numbers of input and output channels to the tile. The output dimensions of each tile are calculated directly from input width and height, filter dimensions, striding, and zero-padding parameters.
4D-tiles have  three main features essential for near-memory acceleration of deep ConvNets:

\textbf{Row-major data layout:} With the conventional tile-oblivious layout, data is fragmented in DRAM, so several DMA transfers are required to fetch one tile. Even a DMA engine with striding capabilities does not help with the inefficiency caused by opening a DRAM row with closed policy \cite{HMCSTANDARD} and partially reading from it in strides.
To address this problem, we modify the underlying storage of the intermediate layers in DRAM to a row-major form (illustrated in \autoref{fig:tile4d}c,e). This way with a single large DMA transfer request, the whole tile can be fetched by the processing cluster. This improves DRAM's read performance which can be exploited as described below. The implications of this mechanism on DMA write and its overheads will be explained later in this section.

\textbf{Tile overlapping:} When the convolution filters are larger than 1$\times$1, borders of the adjacent tiles of each tile should be fetched from DRAM, as well.
Assuming that the bandwidth overhead of these overlapping regions can be tolerated by proper choice of tile dimensions, still, the storage impact on the row-major data placement in DRAM is not trivial, and fragmented DMA transfers will be required to fetch the overlaps.
This problem can be solved by storing the overlapping regions in the DRAM once per each tile. This means storing the ``augmented-tiles'' (shown in \autoref{fig:tile4d}c) instead of ``raw-tiles'' inside DRAM in a row-major form, at the cost of increased DRAM storage and bandwidth. When reading from DRAM, a complete tile (including all overlapping regions required to compute the convolution in its borders) can be fetched using a single DMA transfer request. But, when writing the results back to the DRAM some care should be taken to convert the raw output tile to an augmented-tile for the next layer (explained below).
The average increases in DRAM bandwidth and storage incurred by this mechanism were found to be less than 10\% and 3\%, respectively. Also, on the average around 200\,MB of DRAM was used with maximum usage of 580\,MB for ResNet with 4M-pixel images. %

\textbf{Partial Computations:} Tiling of channels ($T_{Ci}^{(l)}$ and $T_{Co}^{(l)}$)  requires maintaining partial computations, as more than one input tile contributes to the result of each output tile. Assuming that one input tile and one output tile can fit in each cluster's SPM, we perform the following steps to compute each output tile:
Tile $M$  (See \autoref{fig:tile4d}d) and the related filter coefficients ($K_{MQ}$) are fetched from the DRAM. Then, $Q = Q + M*K_{MQ}$ is computed inside the SPM ($Q$ containing partial sums of the output channels). Next, Tile $N$ and $K_{NQ}$ are fetched from the DRAM, and $Q = Q + N*K_{NQ}$ is computed, and so forth. After all input tiles have been read once, activation and pooling are directly performed on the output tile $Q$ (again inside the SPM) and then $Q$ is written back to the DRAM by the associated PE. This mechanism reduces DRAM's write bandwidth and puts more pressure on read bandwidth given that data is only written back once after several DRAM reads (as described), after reduction operations (pooling, strided convolution) which further reduce the number of DRAM writes in comparison with DRAM reads.
In all experiments of this paper, DRAM's write bandwidth was found to be less than 4\% of the read bandwidth. This suits our row-major data layout, requiring DRAM writes to be off the execution critical path.

It is important to explain how the raw output tile of one layer ($l$) is converted to an augmented tile for the next layer ($l+1$), given that data cannot be ``magically'' reorganized in the DRAM. Looking at $T_0$ in \autoref{fig:tile4d}e, we can see that it has 4 regions ($raw$, $A$, $B$, $C$). The $raw$ region of $T_0^{l+1}$ is written to DRAM using multiple fragmented DMA writes when $T_0^{l}$ is computed in SPM. This is shown in \autoref{fig:tile4d}f. The $A$, $B$, and $C$ regions of $T_0^{l+1}$ are written to DRAM after $T_1^{l}$, $T_3^{l}$, and $T_4^{l}$ are computed, respectively, using small DMA chunks shown in \autoref{fig:tile4d}f. Zero-padding is also properly handled at this stage for the corner tiles. Since DRAM writes are off the critical path, we can afford to perform these conversions, without incurring significant overheads.
Another key point is that the raw-tile width and height of the consecutive layers must be equal (for consistent row-major data layout) unless there has been a strided convolution \cite{DLBOOK} or pooling stage between them, for which the tile dimensions will shrink. This way, as we move forward through the ConvNet layers, tile width and height ($T_{Xi}^{(l)}$, $T_{Yi}^{(l)}$) tend to shrink. To avoid this having a negative impact on computation and SPM usage efficiency, we need to increase $T_{Co}^{(l)}$ or $T_{Ci}^{(l)}$. This completely modifies the shape and number of the tiles in each layer and impacts everything from synchronization overheads to the efficiency of the computing loops and DRAM bandwidth.
This highlights the need for a flexible computing cluster to support a wide range of tile dimensions.

\subsection{Mapping Tiles to Clusters} \label{mapping}

Since there is no data overlap among augmented-tiles (except possibly for some filter coefficients), each cluster can execute one tile at a time. This minimizes communication among the clusters. Also, tiling information is prepared off-line (only once) and is stored in a list accessible by all clusters in DRAM.
The master PE (the first PE in each cluster) consults this list to obtain the required information (e.g. address in DRAM, size, and filter coefficients) for the next tile. Then it issues a DMA read to fetch the new tile.
Each cluster works based on ping-pong buffering to hide the setup and DMA transfer latencies. While one tile is being computed by the NSTs in the cluster, another tile is fetched by the master PE and tiling information is prepared for it.
This procedure continues until all tiles in a layer are finished. At this point, all clusters are synchronized before proceeding with the next layer.

Inside each cluster the master PE partitions the tile among the NSTs in the order of $T_{Xo}^{(l)}$, $T_{Yo}^{(l)}$, and $T_{Co}^{(l)}$ dimensions first. This is to ensure that each output is written exactly by one NST, and to remove synchronization requirements among the NSTs.
If still more NSTs are remaining (e.g. for small corner tiles), $T_{Ci}^{(l)}$ is used for tile partitioning, posing some synchronization overheads to the PEs. Therefore, corner tiles (with smaller dimensions) and arbitrarily sized-tiles are properly handled in this scheme.
Thanks to this tile-mapping mechanism, NSTs can work independently without worrying about getting synchronized with each other. Any required synchronization is handled by the RISC-V PEs, through hardware primitives devised for this purpose.
Given that ($X_{o} \times Y_{o} \times K_{x}  \times K_{y}  \times C_{i}  \times C_{o}$) MAC operations need to be done in each layer, 4D-tiling can be viewed as a schedule (in time and space) of this computation to the available resources in NeuroCluster.
Overall, the computation inside each SMC is done in a self-contained manner, without synchronizing with the host processors. The user only offloads a ConvNet task to the SMC, and the rest of the computation happens completely inside the cube. The serial-links are turned-off when not required to save energy. The performance and energy advantages of this scheme are studied in \autoref{multismc}.

\begin{figure}[!t]
	\centering
	\includegraphics[width=0.9\columnwidth]{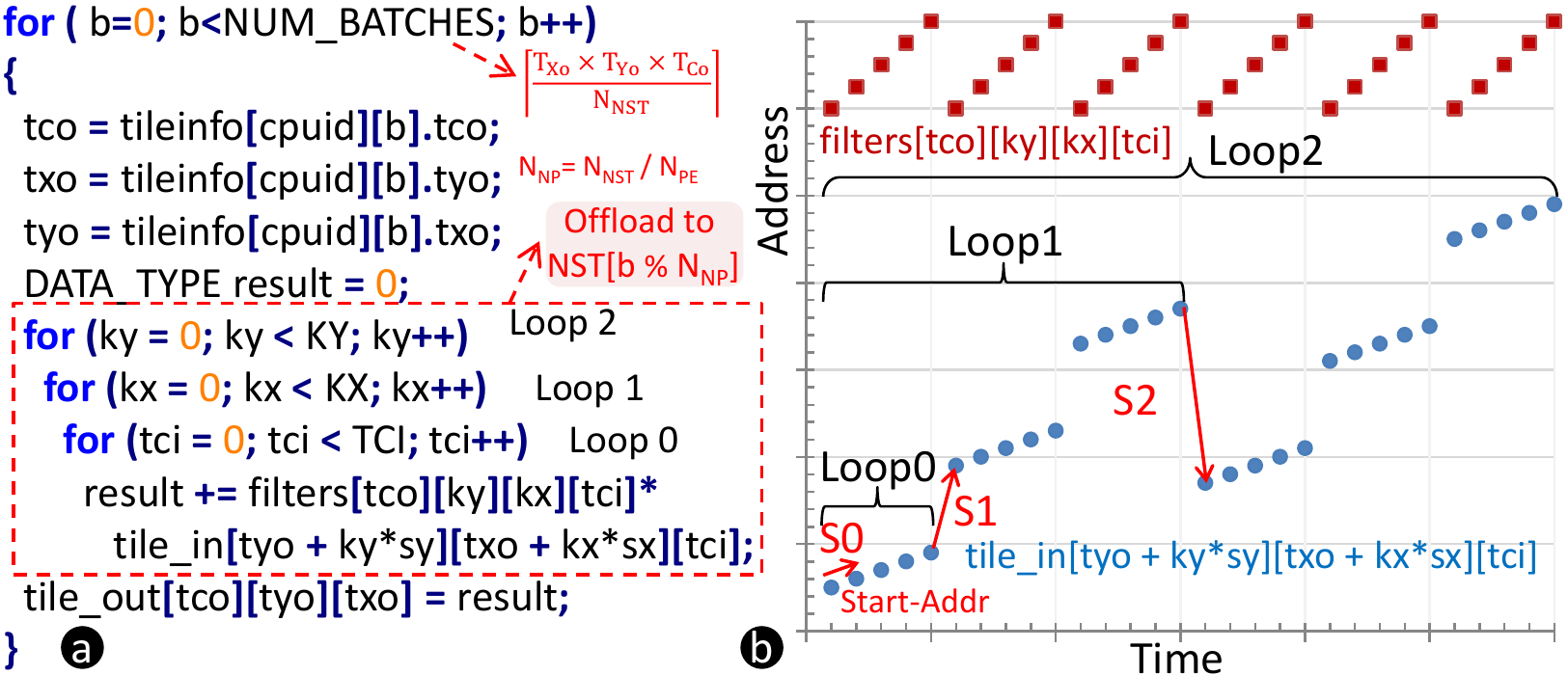}
	\caption{(a) A convolution kernel to be performed on a 4D-tile, (b) a typical memory access pattern generated by this kernel.}
	\label{fig:conv-pattern}
\end{figure}

\section{Programming Model} \label{prog-model}

Previously in \cite{ERFANARCS16}, a complete software stack (API, Device Driver) had been developed for a single-processor PIM device residing on the SMC, exposing it to the user-level applications. This software stack is available online for reuse and modification \cite{SMCSIM}.
An optimized memory virtualization scheme was developed in \cite{ERFANARCS16}, as well, for zero-copy data sharing between host and PIM, allowing PIM to directly access user-space virtual memory without costly memory copies.
In this paper, we have adopted this software stack and extended it to support NeuroCluster, a parallel-processing platform rather than a single core. It has been, also, amended to support DL primitives and offloading of ConvNet tasks. The memory virtualization scheme has been adopted from \cite{ERFANARCS16}, as well.

As a demonstrative example, suppose that the user application wants to execute GoogLeNet on PIM for an image already stored in DRAM. After initializing PIM's API, it uses this API to offload the precompiled computation kernels, including the computing loops for the ConvNet layers (e.g CONV, ACT, and POOL), to NeuroCluster. 
This procedure is done only once. Next, the pointer to the image is passed to the API, and a special table called slice-table (a generalized form of page-table) is built for the data structures, by the driver, and stored in DRAM. The user then triggers the actual execution through the API and waits for the task to complete. The RISC-V cores work directly on the virtual memory and consult the slice-table whenever a miss occurs in their TLB.
The offloading overheads have been previously shown to be negligible in \cite{ERFANARCS16}. Also, in case of having several video frames instead of images, the same pointers can be reused in double/multi-buffering modes to avoid the need for rebuilding the slice-table upon every execution. More details on the software stack and the memory virtualization scheme can be found in \cite{ERFANARCS16}.
\autoref{prog-nst} describes how NSTs are programmed by the PEs to perform the tasks related to inference in ConvNets. \autoref{training} presents the implications of supporting training.

\subsection{Inference with NSTs} \label{prog-nst}

\begin{figure}[!t]
	\centering
	\includegraphics[width=0.9\columnwidth]{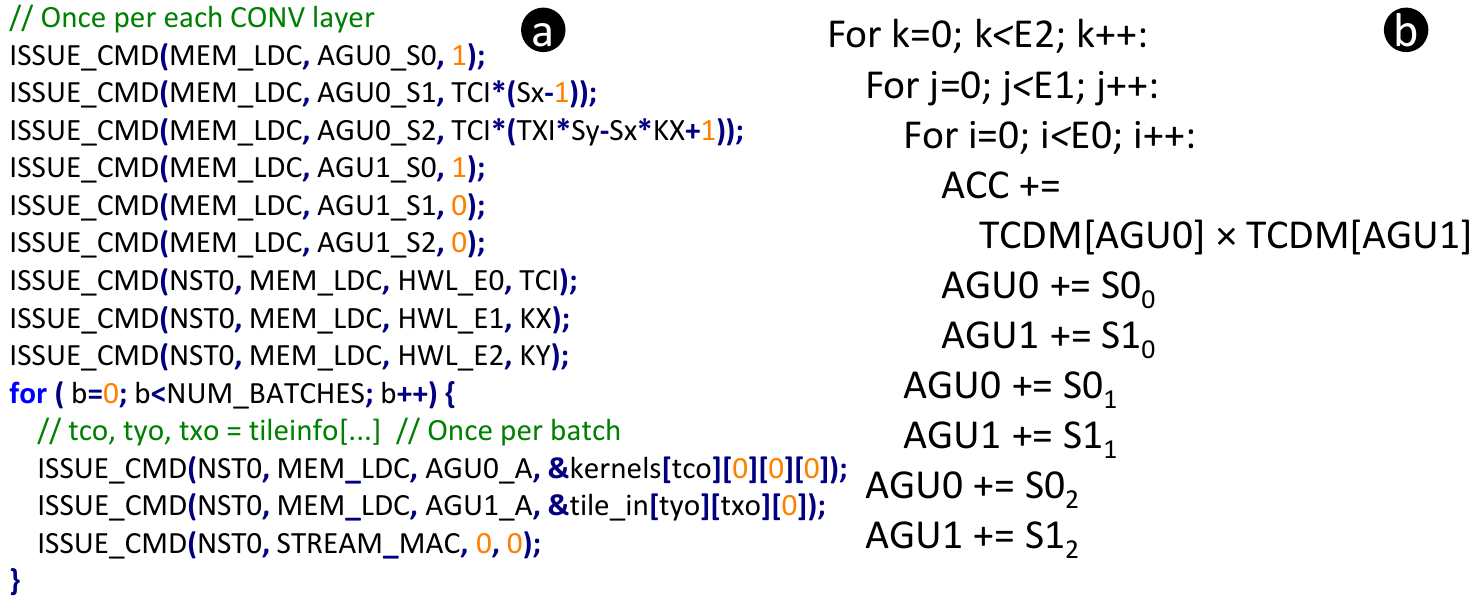}
	\caption{(a) Using NSTs to accelerate the loop shown in \autoref{fig:conv-pattern}a, (b) the pseudo-code for implementation of \textit{STREAM\_MAC} inside the NSTs.}
	\label{fig:stream-mac}
\end{figure}

\autoref{fig:conv-pattern}a illustrates a convolution kernel to be performed on a 4D-tile. The data-structure \textit{tileinfo} contains the required partitioning information for the given tile among the NSTs. When the number of total jobs ($T_{Xo} \times T_{Yo} \times T_{Co}$) is more than $N_{NST}$, the jobs will be broken into several batches (\textit{NUM\_BATCHES}). The flexibility provided by \textit{tileinfo} allows us to reduce the number of convolution loops down to 4 instead of 6. \textit{filters} and \textit{tile\_in} are the two data structures accessed in every iteration of the inner-loop. Typical memory access patterns for this kernel are plotted in \autoref{fig:conv-pattern}b. These patterns seem fairly regular, therefore, NSTs should be easily able to generate them, as well.
It is enough to program the configurations registers of an NST with the starting address and the three step values illustrated in \autoref{fig:conv-pattern}b ($S_0$, $S_1$, and $S_2$), and then issue a \textit{STREAM\_MAC} command to it. This way, the three inner loops of \autoref{fig:conv-pattern}a can be replaced by execution in hardware. This is illustrated in \autoref{fig:stream-mac}a. The latency overheads of these commands are hidden by having multiple NSTs and by filling up their command queues with multiple commands. %
The implementation of \textit{STREAM\_MAC} inside the NSTs is depicted in \autoref{fig:stream-mac}b. This is hard-coded in the main controller of the NSTs and is executed efficiently, without losing any cycles (See \autoref{perf} for results).

Similarly, each NST is able to perform ReLU activation on arbitrary tiles using \textit{STREAM\_MAX} command devised for this purpose on the same set of state machines and hardware blocks. For the sake of generality, \textit{STREAM\_SUM}, \textit{STREAM\_SCALE}, \textit{STREAM\_SHIFT}, and \textit{STREAM\_MIN} are implemented, as well.
Another widely used operation in ConvNets is pooling \cite{CONVNET-TAXONOMY}. NST supports max-pooling \cite{POOLING} through the \textit{STREAM\_MAXPL} command. Thanks to the flexibility of the AGUs and HWLs, arbitrary tiles with different strides are supported. %
Finally, FC layers can also be implemented using a set of STREAM\_MAC commands, similar to the CONV layers. The CLASS layer, however, is executed on the PEs in the current implementation using the \textit{SoftFloat} library \cite{SOFTFLOAT}.

\subsection{Implications of Training} \label{training}

Backpropagation is the prevalent method for training NNs including ConvNets \cite{DLBOOK}. Given a set of training sample inputs, first a forward propagation is executed layer-by-layer, then using an optimization algorithm, such as gradient descent (GD) \cite{CONVNET-TAXONOMY}, the coefficients (weights) are updated backwards so that the network learns that sample. A modern training algorithm based on GD has three phases \cite{DLBOOK}: (1) Forward Pass, (2) Gradient calculation and routing, and (3) Weight update.
In step (1), the selected input (e.g. an image) is fed to the network and the outputs of all layers including the value of the loss function are calculated.
This is similar to a normal inference pass, except that additional information about the current operating point (e.g., max-pool decisions) in all layers has to be stored, such that it can be retrieved later on for gradient calculation.
This can be easily handled by our platform because plenty of DRAM is available to the NeuroClusters through a high-bandwidth and low-latency 3D interface. For example, ResNet-152 requires 211\,MB for its coefficient and a total of 161\,MB for all its layers. This aggregates to a total of 372\,MB of DRAM storage. %
Another difference with inference is that the POOL layer should keep track of the inputs which were maximal in the pooling operation. This is called \textit{argmax} and since just comparison with zero is involved, in this implementation we use the RISC-V cores for it.

In step (2), starting from the final stage of the network (classification layer), the gradients of the loss function are calculated with respect to the inputs ($D_X$) and to the weights ($D_W$) and propagated backwards towards the input layers. For the FC layers
$D_X = W^T . D_Y$ and $D_W = D_X . X^T$, and for the CONV layers $D_X = D_Y * W^T$ and $D_W = X * {D_Y}^T$ can be completely calculated on the NSTs using a series of STREAM\_MAC operations ($Y$ is the output gradient of each layer which is propagated backward to the input $X$, and $T$ stands for matrix transpose).
ACT layer only propagates back the gradients ($D_{Xi} = Xi \geq 0 ?\,D_{Yi} : 0$). This operation is not currently supported by NSTs. But since only comparisons with the zero are involved, the integer datapath of RISC-V cores is used. POOL layer, similarly, propagates back the gradients with a matrix scatter operation \cite{DLBOOK}, populating a sparse matrix without performing any actual computation. Again, this operation is implemented on the RISC-V cores in the current version.
SoftMax (CLASS) is calculated similarly to the forward-pass on the RISC-V cores.
Finally, in step (3), the weights are updated either with fixed or adaptive step sizes ($\alpha$ or $\alpha_i$, respectively): $W_i = W_i - \alpha(dW_i + \lambda W_i)$. This procedure is repeated in an iterative manner for all variations of GD algorithms (e.g. Stochastic GD, Batch GD) \cite{DLBOOK}. A fixed step implementation of this formula is currently supported by the NSTs, while adaptive steps need to be calculated by the PEs, once per each backward pass. An estimation for the performance of training on SMC is presented in \autoref{perf}.

\section{Experimental Results} \label{exp}

Our baseline system is composed of a memory-centric network \cite{MEMCENTRIC} of four SMC devices based on a mesh topology. Each SMC hosts a NeuroCluster with 16 clusters on its LoB die, with each cluster having 4 RISC-V cores (with 1kB private instruction cache each), 8 NSTs, a DMA engine, and 128kB of SPM. This configuration is found to achieve reasonable performance and efficiency through several simulations. Total available DRAM is 1GB in 4 stacked dies with DRAM banks of 32MB and a closed-page policy \cite{ERFANTVLSI16}. Low-interleaved-addressing is implemented as the HMC's default addressing scheme \cite{HMCSTANDARD}. A summary of these parameters is also listed on page \pageref{fig:sch}.
A fully functional and cycle-accurate (CA) RTL model of the NeuroCluster has been modeled in SystemVerilog, with the components adopted and reconfigured from \cite{PULP}.
This model along with a previously developed cycle-accurate model of the SMC \cite{ERFANTVLSI16} allows us to analyze the performance of tiled execution over a single SMC device considering the programming overheads.

Silicon area and power consumption are also extracted from these models using topographical logic synthesis (See \autoref{area-power}). %
In addition, an epoch-based in-house simulator is developed (modeling the SMC network shown on page \pageref{fig:sch}) to estimate the performance and power consumption of executing full ConvNets on large images, based on the data obtained from the CA simulations.
This strategy allows us to obtain both reasonable accuracy and very high simulation speed.
Our simulation platform supports CAFFE \cite{CAFFE} representation of the SoA ConvNets. For every layer of the ConvNets under study, the optimum tile dimensions are found based on performance, energy efficiency, available SPM size, and required DRAM bandwidth. This procedure requires multiple simulations with different combinations of parameters and is only done once per each ConvNet (at the beginning). Optimally sized tiles can then be used in later simulations with different images.
Four serial link controllers in LoB are modeled to consume up to 10\,W of power for highest traffic pressure \cite{ERFANARCS16}\cite{HMC}. We can share this 10\,W power budget between the serial link controllers and NeuroCluster, and for example by turning off one of them we give a 2.5\,W power budget to NeuroCluster allowing it to operate in the ``shadow'' of a powered-down serial link.
Performance is studied in \autoref{perf}. Detailed energy consumption and silicon area results are presented in \autoref{area-power}. Finally, the overall results of the multi-SMC network are presented in \autoref{multismc}.

\begin{figure}[!t]
	\centering
	\includegraphics[width=0.6\columnwidth]{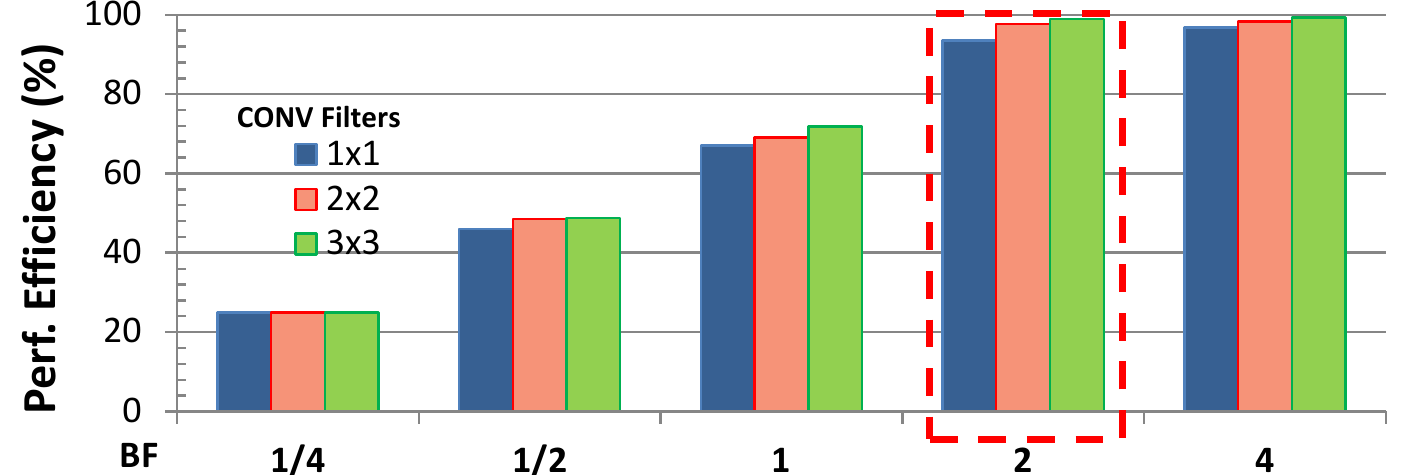}
	\caption{The effect of SPM's banking-factor on the performance efficiency of a single cluster, executing tiled convolutions with 1x1, 2x2, and 3x3 filters over average tiles of the studied ConvNets.}
	\label{fig:g2}
\end{figure}

\subsection{Performance of Single SMC} \label{perf}

\begin{figure}[!t]
	\centering
	\includegraphics[width=1\columnwidth]{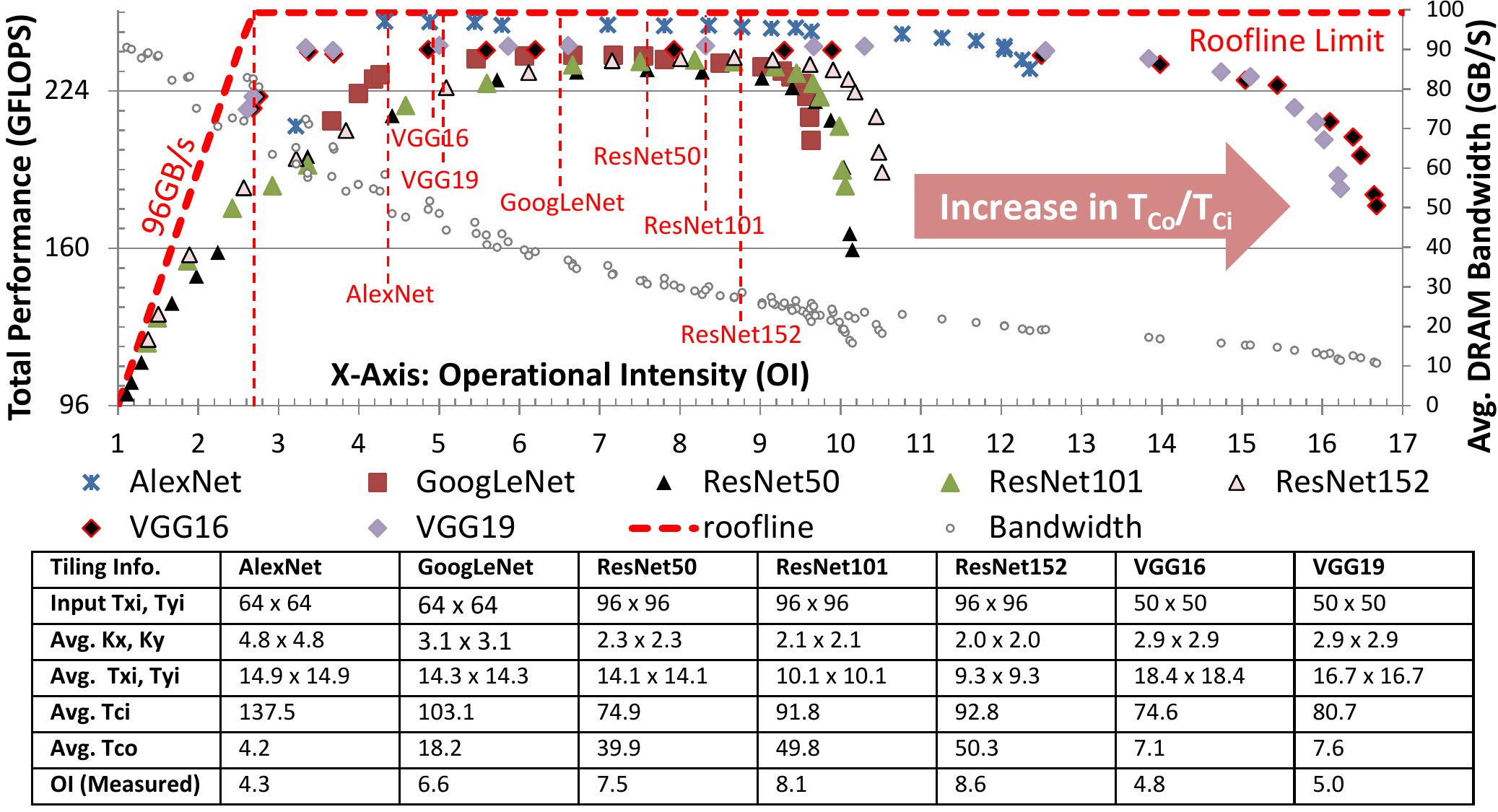}
	\caption{Roofline plot for execution of ConvNets over a single SMC, %
			and the actual tile dimension statistics for different ConvNets.}
	\label{fig:ht10}
\end{figure}

The average performance efficiency (actual/peak performance) of a single cluster measured in CA simulation is illustrated in \autoref{fig:g2}, where the cluster is executing tiled convolution on its NSTs with 1x1, 2x2, and 3x3 filters over tiles with average dimensions of the studied ConvNets, listed in \autoref{fig:ht10}.
We define performance efficiency ($PEF$) as follows:
\begin{small} 
	\[ PEF = \dfrac{Total \# MACs}{\#Cycles \times N_{NST}} \%  \]
\end{small} 

$Total \# MACs$ indicates the total number of MACs performed by all NSTs, and $\#Cycles$ stands for the total number of execution cycles. $PEF$ is an indication for how well and efficiently the NSTs have been utilized.
The banking-factor\footnote{Banking-factor is the ratio between the number of SPM banks and the number of master ports (from the NSTs). 
In WLI memories, this parameter directly affects the ratio of bank-conflicts inside the SPM, and has a critical impact on the clock frequency and area of the cluster interconnect. More information: \cite{ERFANIET}.} (BF) of the SPM is changed from 1/4 to 4 (i.e. from 4 to 64 banks). On the average, BF=2 yields an efficiency of over 93\% for the execution of a single tile. This is why the baseline clusters shown in \autoref{fig:sch} have 32 SPM banks each, in a WLI organization.
Another point is that a traditional bank-level interleaved (BLI) SPM needs to be explicitly managed and partitioned by software, and its performance is highly dependent on the tile dimensions. A significant amount of bank-conflicts can occur if it is not properly managed. Also, with BLI, only half of the banks will be used for computation (because of ping/pong buffering), further reducing the bandwidth.
In this paper, we use WLI because of its flexibility and high parallelism regardless of the tile dimensions. %
One last point to observe in \autoref{fig:g2} is that the execution efficiency reduces as the convolution filters shrink (3x3\,\textgreater\,2x2\,\textgreater\,1x1). %
This is a common trend in modern ConvNets and will be investigated later in this section.

\autoref{fig:ht10} illustrates the roofline plot \cite{ROOFLINE} for complete execution of the ConvNets listed in \autoref{tbl:storage} on a single SMC, along with the tile dimension statistics, where OI denotes operational intensity (previously defined in \autoref{related-conv}).
For each ConvNet, different OI ratios are achieved by altering the tile channel ratios of all layers ($R_{TCL}$ = $T_{Co}^{(l)}$/$T_{Ci}^{(l)}$, directly proportional to OI). In this experiment, $T_X$ and $T_Y$ of the channels are kept constant.
The left axis shows achievable GFLOPS, and right axis shows the average DRAM bandwidth.
The synthetic extensions to ResNet (250K$\sim$1M) have been omitted from this plot as they behaved similarly to the rest of the ResNet group.
This plot highlights the importance of proper tile sizing on the delivered performance, as different ConvNets have different optimal points. Also, too much increase in $R_{TCL}$ can negatively impact performance, because the initialization overhead of the NSTs is highly dependent on this parameter, especially for ResNet with a high percentage of 1x1 filters (as later explained in this section).
We, also, observed that the bandwidth demand varies widely across different ConvNet layers. 
For this reason, we have dedicated 3 high-performance AXI ports (each delivering 32GB/sec) to connect the cluster to the main SMC interconnect (See \autoref{fig:sch}). This is to smoothly adapt to different bandwidth demands with minimal impact on the delivered performance.

\begin{figure}[!t]
	\centering
	\includegraphics[width=0.95\columnwidth]{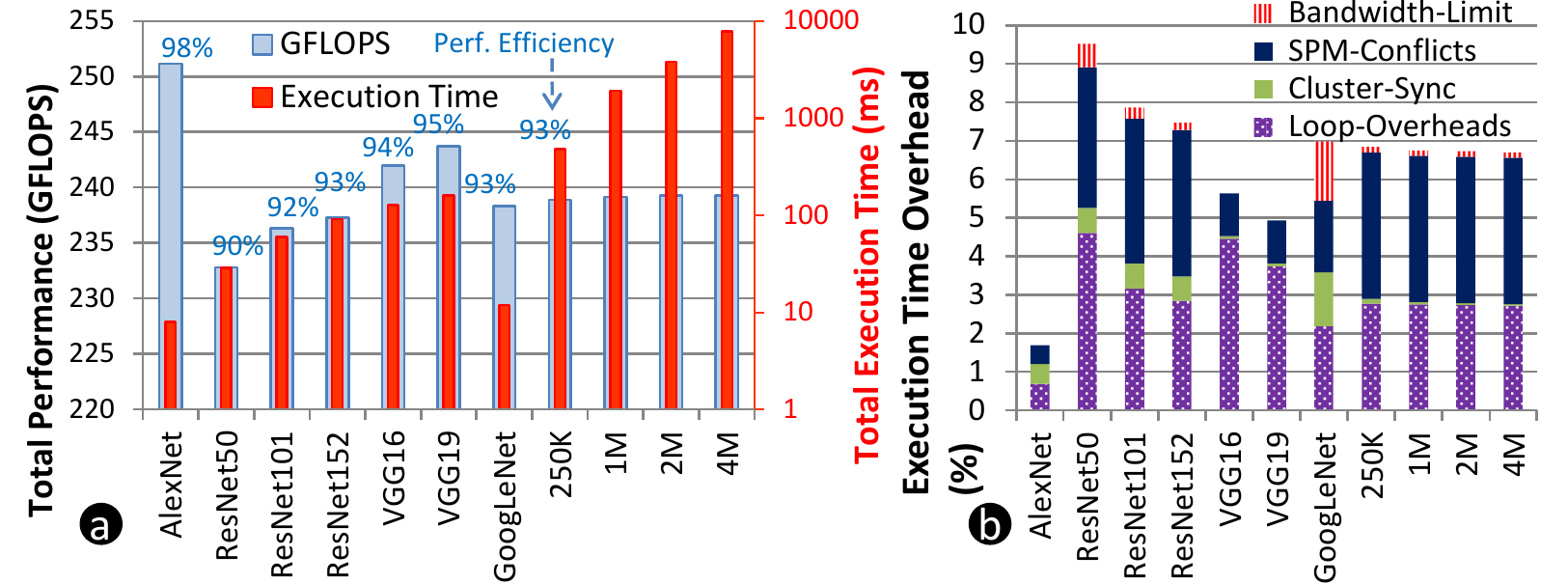}
	\caption{(a) Performance comparison of different ConvNets on a single SMC device, and (b) breakdown of different overheads contributing to performance loss.}
	\label{fig:ht7-8}
\end{figure}

Having found the optimum tile dimensions, \autoref{fig:ht7-8}a depicts the overall performance (GFLOPS) and total execution time for the studied ConvNets on a single cube. Among the studied ConvNets from the SoA, VGG networks have the highest execution time due to their higher requirement for MAC operations, while GoogLeNet and AlexNet are the fastest ones executing in less than 12\,ms. GoogLeNet has a very low computation requirement (less than 2\,GMAC for each forward pass) compared to the other modern networks (ResNet and VGG) mainly due to the use of strided convolution in the beginning layers.
It can be further seen that the ResNet group achieves the lowest performance efficiency. This can be associated with higher percentage of SPM conflicts illustrated in \autoref{fig:ht7-8}b.
In this figure, four main sources of performance loss are identified as: $T_L$: Loop overheads, $T_S$: Cluster Synchronization, $T_B$: Bandwidth limit, $T_C$: SPM Conflicts.
In the NeuroCluster architecture, the RISC-V cores are responsible for tile preparation, loop initialization, DMA setup, and synchronization with other clusters. While NSTs are responsible for the actual computation on the SPM ($T_U$: Useful Computation). For this reason, the RISC-V cores account for ($T_L + T_S$) overhead cycles, while the NSTs account for ($T_C + T_B + T_U$) from the total execution time. Overall, for all studied ConvNets, less than 6\% of the total execution time was spent on the RISC-V cores, and the rest was spent on NSTs, either for useful computation, or waiting for the memory system.  Tile preparation and DMA setup phases are also handled by the RISC-V PEs, nevertheless, they are overlapped with the execution on NST so they are hidden and do not contribute to the total execution time.
It is also worthy to note that among the overheads shown in \autoref{fig:ht7-8}b, only loop overheads are caused by the proposed tiling mechanism, which account for less than 5\% of the performance loss.

\begin{figure}[!t]
	\centering
	\includegraphics[width=0.9\columnwidth]{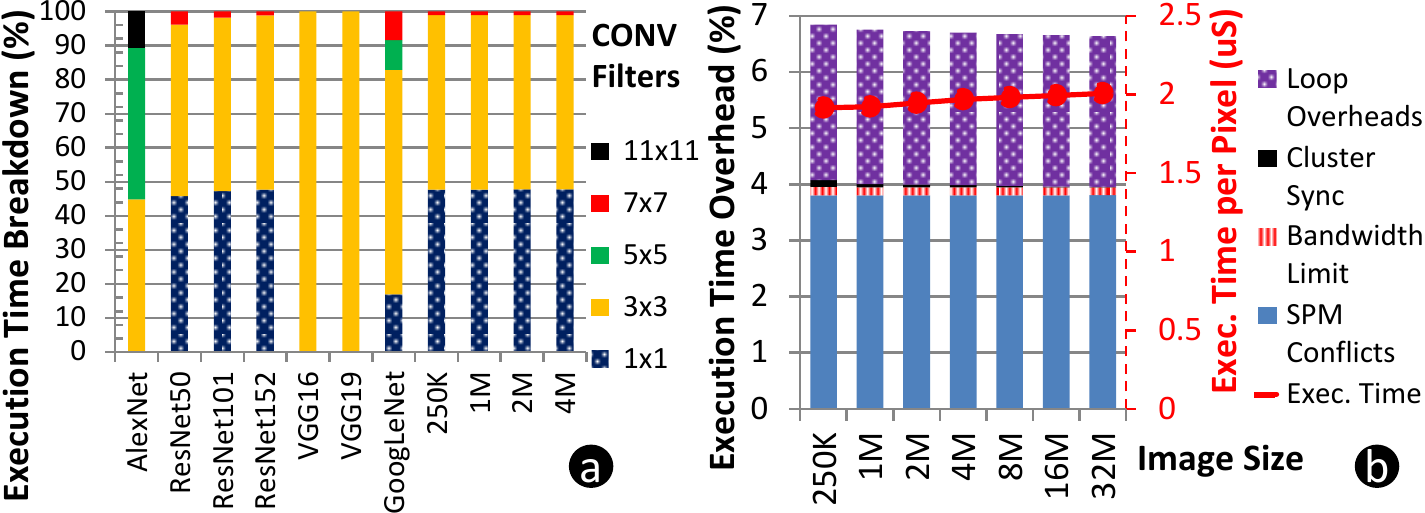}
	\caption{(a) Breakdown of total execution time versus the size of the convolution filters. (b) Execution overheads (left-axis), and execution-time per pixel versus image size for ResNet-based networks (right-axis).}
	\label{fig:ht5ht11}
\end{figure}

To gain further insights from this plot, a breakdown of total execution time is depicted in \autoref{fig:ht5ht11}a versus the size of the convolution filters. %
As can be seen, a significant portion of the execution time (over 45\%) of the ResNet group is spent in 1x1 filters, and as we saw previously in \autoref{fig:g2}, 1x1 filters cause more SPM conflicts than larger filters. Plus, for 1x1 filters the 3 convolution loops illustrated in \autoref{fig:conv-pattern}a change into a single loop iterating over $Tc_i$. This increases the relative overhead of NST initialization (shown in \autoref{fig:ht7-8}b).

To demonstrate the scalability of the proposed PIM platform, the size of the input images is increased from 250K-pixels to 32M-pixels, and execution-time per pixel with execution overheads for ResNet-based synthetic networks are plotted in \autoref{fig:ht5ht11}b.
This plot clearly shows that the execution overheads do not increase even for very large images, and execution-time per pixel only increases slightly due to the increased number of layers.
This proves the effectiveness of the proposed 4D-tiling mechanism and the efficient computation paradigm based on ping-pong buffering to hide the latencies. %
To summarize each SMC instance is capable of processing 126, 83, 34, 16, 11, 8, 6 frames (each frame being 220$\times$220$\times$3 Bytes) per second for AlexNet, GoogLeNet, ResNet50, ResNet101, ResNet152, VGG16, VGG19, respectively, with an average performance of 240\,GFLOPS. This performance is scalable to larger images as described.

Lastly, an estimation of the training performance on SMC can be obtained by answering two questions: (I) How much additional computation do each of the training stages need compared to their corresponding stage in inference? (II) How much efficiency is lost due to the extra usage of the RISC-V cores for computation?
For the forward pass, the only extra operation is the \textit{argmax} function in the POOL layer. For gradient routing, FC and CONV layers require additional computations (on the NSTs), while ACT and CLASS need the same amount as inference. POOL, also, implements a different operation (on the RISC-V cores) as described in \autoref{training}. Finally, the weight-update phase works solely on the FC and CONV layers, and its overhead is found to be less than 5\% of the total execution time.
We have chosen GoogLeNet as a representative of the future ConvNets with strided convolutions, shrinking kernel coefficients, and complex topologies \cite{DLBOOK}.
For GoogLeNet we have estimated the execution-time of each kernel relative to its corresponding kernel in inference. We have, then, scaled the execution times of inference with these results. \autoref{tbl:training} summarizes these estimates, where ``Training (Best)'' indicates the estimated execution-time provided that the NSTs implement the additional required functions such as \textit{argmax} and vector multiply. This is not achievable in the current version, and it is planned to be done as a future work.
``Training (Current)'' is the estimated execution time with the current platform. It can be seen that one training pass takes 3.6X longer than one inference pass (almost 3X more in the best case).
 This is reasonable and consistent with our previous measurements on GPUs \cite{LUKAS-DAC15}. Also, the amount of efficiency loss due to using RISC-Vs for part of the computation is 17\%. Overall, the training performance can be estimated as 197\,GFLOPS for GoogLeNet.

\begin{table}[!t]
	\centering
	\caption{Execution-time of training (ms) compared with inference for different layers of GoogLeNet.}
	\includegraphics[width=0.75\columnwidth]{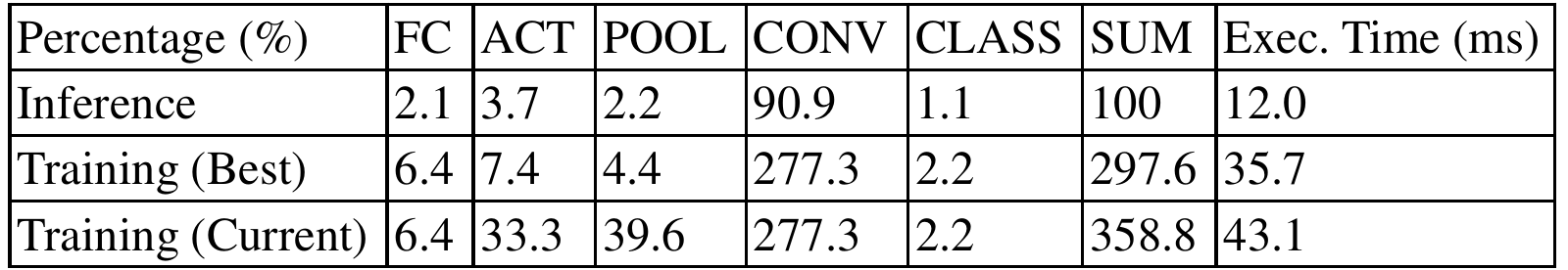}
	\label{tbl:training}
\end{table}

\subsection{Silicon Area and Power Efficiency} \label{area-power}

\begin{figure}[!t]
	\centering
	\includegraphics[width=0.9\columnwidth]{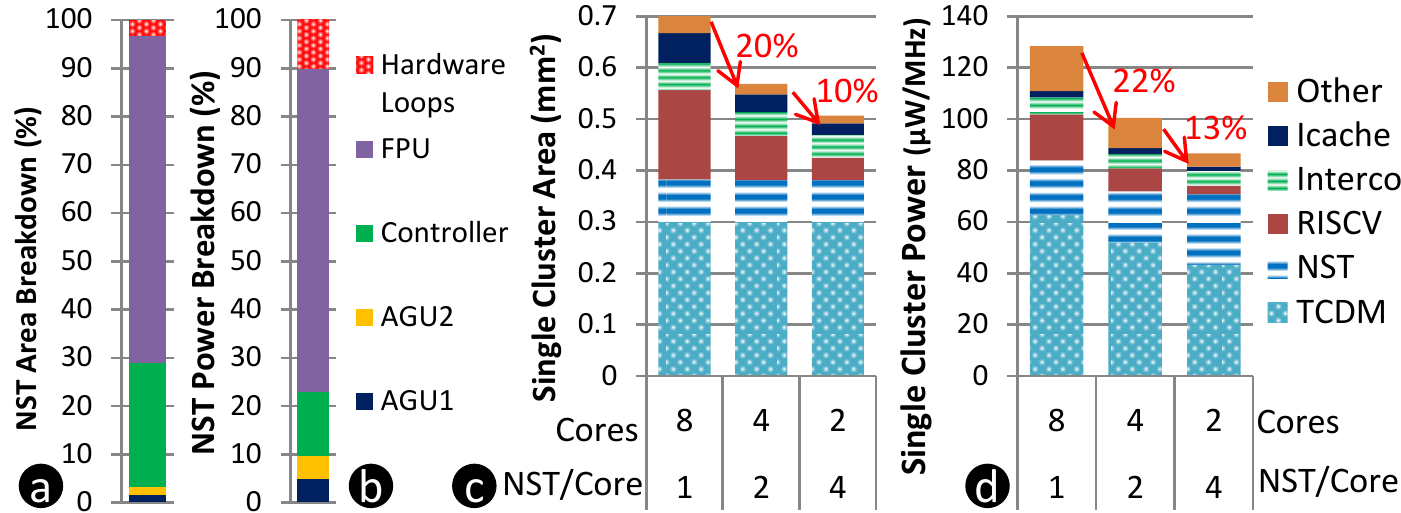}
	\caption{(a) Area and (b) power breakdown inside one instance of NeuroStream, (c) Area and (d) power breakdown for one of the clusters shown in \autoref{fig:sch}, where the number of PEs has been changed from 2 to 8.}
	\label{fig:g1-nst-cluster}
\end{figure}

A complete processing cluster was synthesized using Synopsys Design Compiler (J-2014.09-SP4) in the topographical mode in 28nm FDSOI technology of STMicroelectronics (1.0$V$, SS, 125$^{\circ}C$, Low Threshold Voltage Transistors), achieving a clock frequency of 1\,GHz.
The critical-path was inside the NST blocks where MAC is computed. Our current implementation uses discrete DesignWare IP components in IEEE compatible mode. This creates a timing critical loop which cannot be pipelined. As a future work, we plan to switch to fused carry-save MAC with a fast carry-save adder in the loop.
Power consumption was extracted using Synopsys Primetime (H-2013.06-SP2) at 25$^{\circ}C$, TT, 1.0$V$. The CA cluster model runs the tiled-convolution illustrated in \autoref{fig:conv-pattern}a on typical tiles (listed in \autoref{fig:ht10}) by offloading them to NSTs similar to \autoref{fig:stream-mac}a. Switching activity is then recorded and fed to Primetime for power extraction.
For vault controllers and the SMC controller previously developed models in \cite{ERFANTVLSI16} and \cite{ERFANARCS16} were used (all in 28nm FDSOI), the serial link area and energy were estimated based on \cite{HMC}\cite{SERDES}.
5000 TSVs \cite{EXASCALE} with a pitch of 48 $\mu m$  $\times$ 55$\mu m$ \cite{HBM} were used to estimate TSV matrix area, with energy modeled from \cite{HMC}.

\begin{figure}[!t]
	\centering
	\includegraphics[width=0.95\columnwidth]{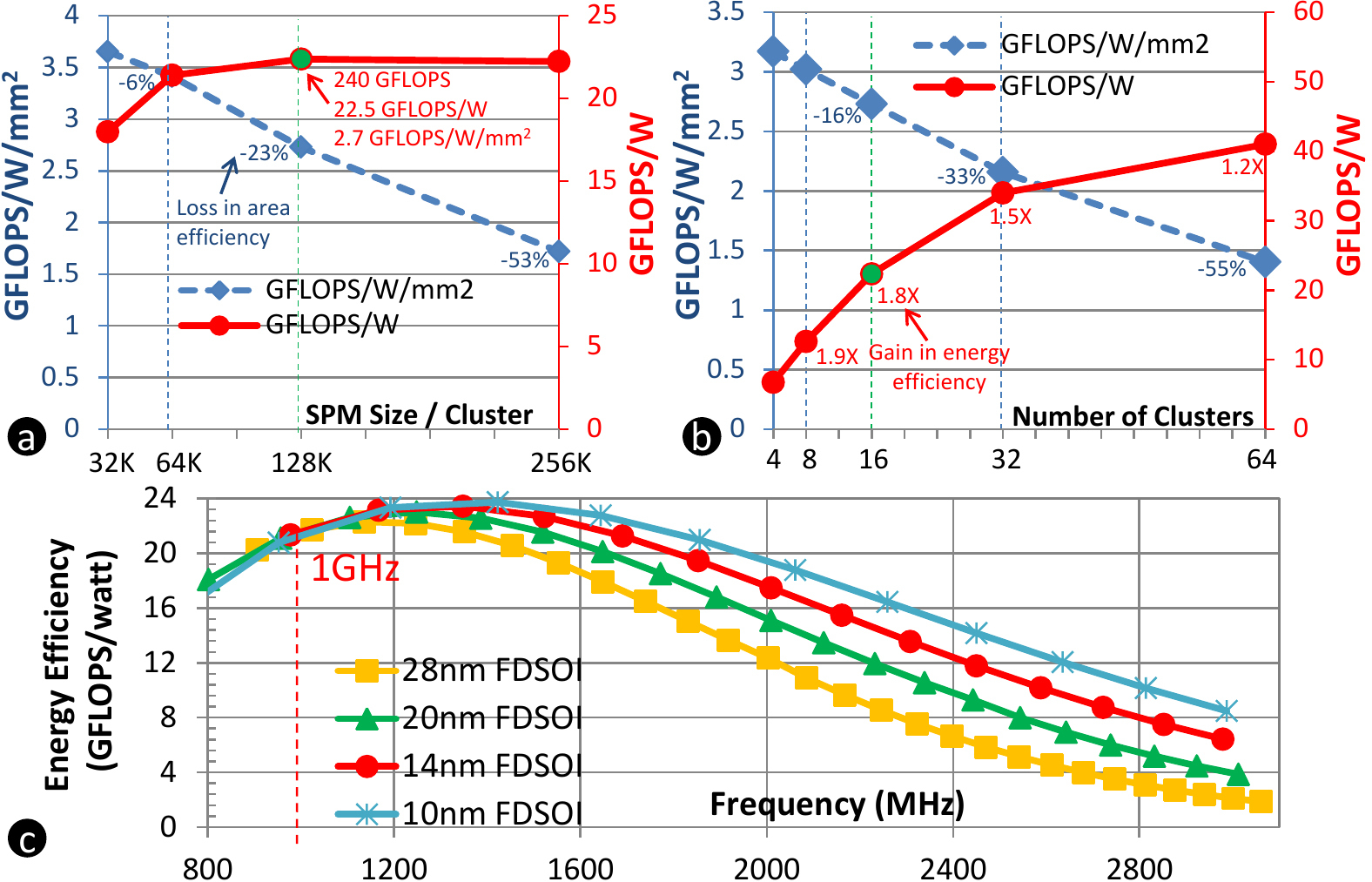}
	\caption{(a) Effect of the SPM size per cluster (b) and the number of total clusters, on energy and area efficiency of the SMC. (c) Effect of technology scaling of the LoB die on the energy efficiency of a complete SMC device. Voltage is scaled to achieve different operating frequencies.}
	\label{fig:ht131918}
\end{figure}

\autoref{fig:g1-nst-cluster}a,b illustrates the power and area breakdown inside one NST instance.
As expected, over two-third of the total power and area are dedicated to the streaming FPUs,
while the controller accounts only for 26\% of the total area and 13\% of the power consumption. 
These simplified FSMs allow for a significant energy reduction compared to full-featured processors such as RISC-V (see below).
\autoref{fig:g1-nst-cluster}c,d illustrates the area and power breakdowns for a complete cluster, where the number of PEs is changed. Decreasing the number of RISC-V core from 8 to 4 gives an average power and area reduction of 20\% and 22\%, respectively, while from 4 to 2 smaller reductions of 10\% and 13\% are obtained.
Reminding that having fewer RISC-V cores increases the programming overheads (each core needs to manage more NSTs), we choose to have 4 cores per cluster. For the NSTs, 8 instances were found optimal, as beyond that the delay and area of the cluster-interconnect limits achievable frequency.
The optimal SPM size in terms of energy-efficiency was found to be 128kB per cluster. This is shown in \autoref{fig:ht131918}a, yielding 22.5 GFLOPS/W and 2.7\,GFLOPS/W/mm$^2$.
Thanks to the proposed tiling scheme, this choice does not affect the size of the supported ConvNets and images, and very large networks like \textit{4M} can be easily executed on this platform, regardless of the SPM size, as long as they are  properly tiled.
Also, we found that deploying 16 clusters leads to a well-balanced design as shown in \autoref{fig:ht131918}b. More clusters negatively impact the area-efficiency, yields diminishing energy returns, and can cause major modifications in the standard HMC's stack structure. This is explained below at the end of this section.

The choice of a moderate operating frequency in our design is justified through the experiment shown in \autoref{fig:ht131918}c for different technology choices. The energy efficiency of a complete SMC device executing different ConvNets is estimated for the 28nm to 10nm FDSOI technologies at various operating points.	
Interestingly, a moderate clock frequency of 1.2\,GHz achieves the highest efficiency, and increasing the clock speed beyond that is not beneficial. This is mainly due to the communication bound (DRAM's latency and bandwidth), limiting the achievable performance. This choice, also, relieves us from thermal concerns. As in \cite{NEUROCUBE} a 3D thermal simulation of the HMC with 4 stacked DRAM dies shows that a processing cluster clocked up to 5\,GHz can be embedded in its LoB die increasing the temperature up to 76$^{\circ}C$ which is below the thermal limits of HMC \cite{HMCSTANDARD}.
Power consumption is a secondary concern, as well, because up to 10\,W budget is available in the LoB die by turning-off the serial links, and NeuroCluster only consumes 2.2\,W.

\begin{figure}[!t]
	\centering
	\includegraphics[width=0.88\columnwidth]{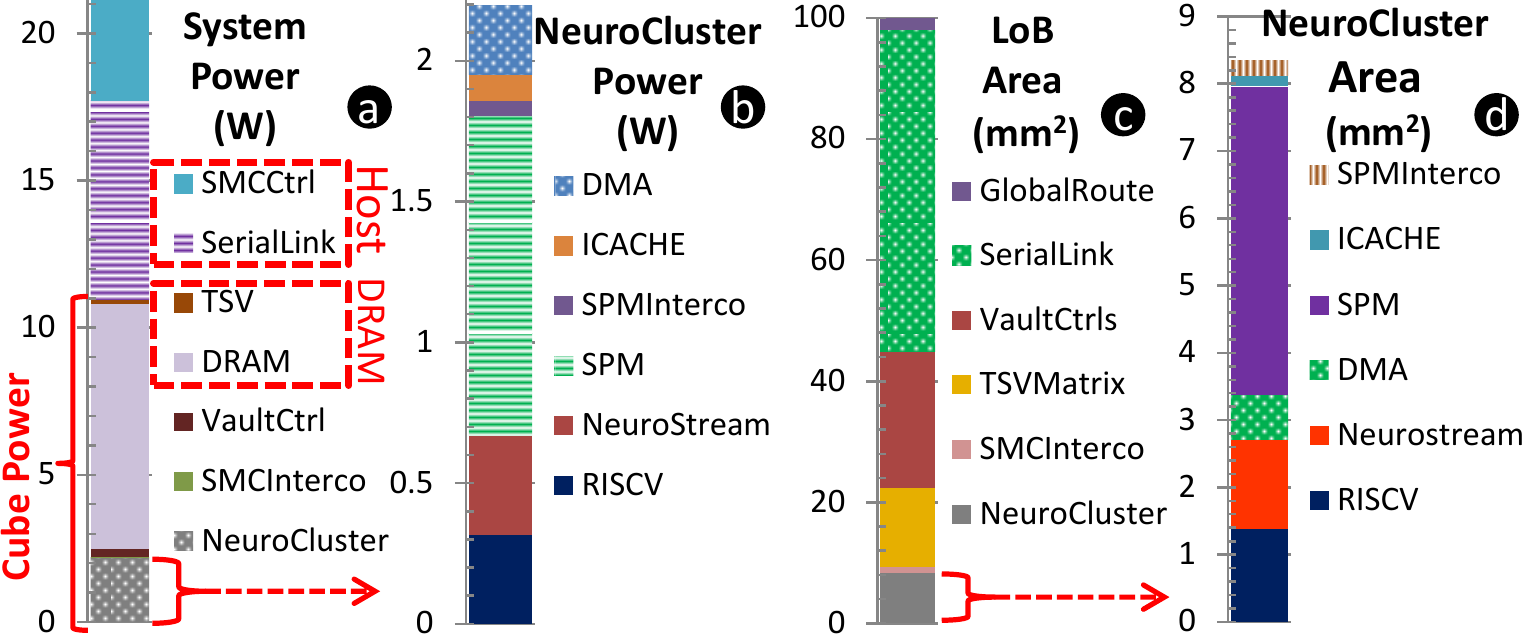}
	\caption{(a) Total system power for NeuroCluster placed on the host side, (b) power breakdown inside NeuroCluster. (c) Silicon area of the whole LoB die and (d) one NeuroCluster.}
	\label{fig:ht34}
\end{figure}

\autoref{fig:ht34}a depicts different contributors to power consumption in the baseline configuration. \textit{Cube Power} represents the total consumed power inside a single SMC averaged for the execution of all ConvNets under study. As can be seen 11\,W is consumed inside the cube, on the average. The NeuroCluster is responsible only for 2.2\,W of this power.
While the rest (around 75\%) is consumed inside the DRAM dies, mostly due to the refresh operations. For this reason, the average power does not vary a lot for the execution of the studied ConvNets (less than 10\%).
The power consumed in NeuroCluster is dominated by the SPM (51\%) and the NSTs (16\%), while the RISC-V cores only consume 14\% (See \autoref{fig:ht34}b). This highlights the efficiency of this architecture, minimizing the unnecessary power consumed in control-flow and dedicating it to the computations.
Each NST instance consumes an average of 2.7\,mW, while a RISC-V core consumes 2.2\,mW just for programming and coordinating the NSTs.
Floating-point arithmetic is only responsible for a small portion (less than 3\%) of the total power in our platform, and any improvements to it (e.g. reduced precision) is expected to have a marginal power reduction in the overall system.
For this reason, we have kept FP32 for generality and flexibility in supporting different workloads and focused on optimizing the rest of the system (especially the DRAM interface).
Overall, an energy efficiency of 22.5\,GFLOPS/W is achieved inside one SMC for the execution of complete ConvNets (NeuroCluster itself achieves 117\,GFLOPS/W). 
One interesting observation is that if we place the NeuroCluster accelerator on the host side (behind the SMC controller and the serial links) rather than inside the SMC, while maintaining exactly the same computation paradigm, the total execution time and the power consumed in the NeuroCluster itself do not change much, but on the average, system power increases by 10.2\,W. This power is consumed in the SMC controller and the serial links, suggesting that computing inside the SMC can give an average energy reduction of 48\% compared to a similar host-side accelerator (1.7X improvement in system-level energy efficiency).
Another downside of the host-side accelerators is that they require more buffering to deal with the higher memory access latency and to maintain a constant bandwidth (Little's law).

Finally, looking at the silicon area results in \autoref{fig:ht34}c,d we can see that the NeuroCluster (8.3 $mm^{2}$) occupies around 8\% of the area in the LoB die with the SPM (55\% of NeuroCluster), the RISC-V cores (16.5\%), and the NSTs (16\%) being its main contributors. The total area for the LoB die was estimated as 99$mm^{2}$.
It is worth mentioning that the LoB die of HMC is already larger than its DRAM dies \cite{HMC}, and it is occupied by four serial link controllers ($\sim55mm^2$), 32 vault controllers ($\sim25 mm^2$), and the TSV matrix ($\sim13mm^2$), with almost no free space available in it. Any major addition to this die requires modification of the 3D stack structure and power delivery network. This is why we have tried to keep the area increase to a maximum of 3\% in each dimension.
The optimal parameters found in this section were listed in \autoref{intro} and used in all experiments.

\subsection{The Multi-SMC Network} \label{multismc}

This section presents the estimated performance and energy efficiency for the SMC network previously shown on page \pageref{fig:sch}.
Four SMC devices are connected to each other using mesh topology with an HD camera recording raw images of 8\,M-pixels.
ResNet has been chosen as the most difficult to accelerate ConvNet among the studied ones due to having a significant percentage of 1x1 and 3x3 kernels (more than 95\%), with a very large number of layers (more than 100).
The camera sends the images to the memory cubes over the highlighted links in \autoref{fig:sch}a, and each SMC executes ResNet on one complete frame, independently from the other cubes. This ensures minimum communication among the cubes and allows for turning off the serial-links for a large portion of the time. Each SMC has a copy of the ConvNet coefficients inside its DRAM dies, and the coefficients have been preloaded once at the beginning.

The host system-on-chip (SoC) is only responsible for coordination and receiving the results. It does not send or receive data at a high bandwidth, yet we keep its serial link (Link0) always active, to make sure it can manage the other devices through that link.
The other serial links, however, are turned on only when there is a data to send over them, and then turned off again.
Considering a typical bandwidth of 16\,GB/sec for each serial link, and the power-state transition times obtained from the HMC specifications V2.1 \cite{HMCSTANDARD} [Active to Sleep: $600ns\,(t_{SME})$,
Sleep to Power-Down: $150 \mu s\,(t_{SD})$,
and Power-Down to Active: $50 \mu s$], the total power consumed in the SMC network can be estimated as 42.8\,W.	
The camera sends images to the cubes in a ping-pong fashion: while an SMC is working on one image, the camera sends another image to its DRAM. This is easily achievable because there is plenty of space available inside each SMC.
Our SMC network can achieve 955 GFLOPS @42.8\,W. %
Moreover, this architecture is scalable to a larger number of nodes because the average bandwidth demand over the serial links is not large (in the order of 10\,MB/sec per image). Therefore it is possible to turn on a link, transfer the image at a higher bandwidth, and turn it off, periodically. This asynchronous and on-demand mechanism allows us to achieve a scalable performance with high energy efficiency.

\begin{figure}[!t]
	\centering
	\includegraphics[width=0.88\columnwidth]{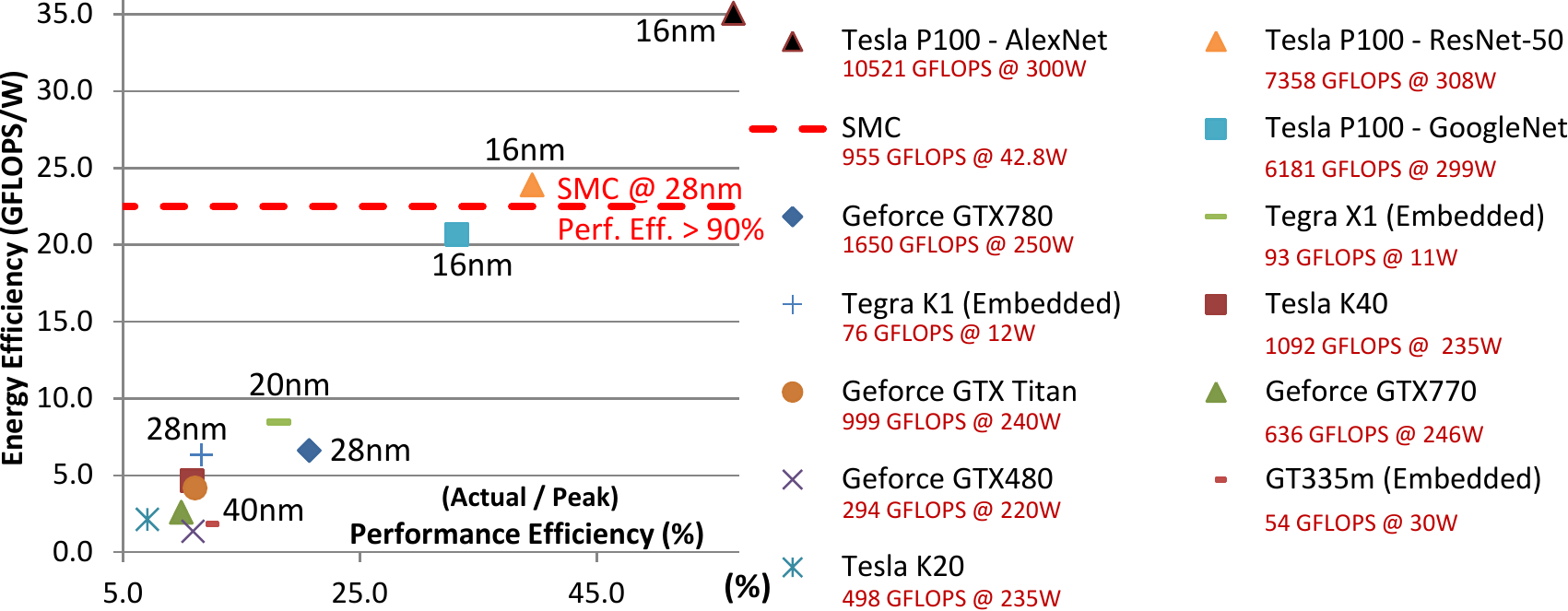}
	\caption{Energy and performance efficiency of the SoA ConvNet implementations on GPUs with standard frameworks.}
	\label{fig:ht56}
\end{figure}

To extend our comparison with the SoA GPUs, \autoref{fig:ht56} presents the energy and performance efficiency of some of the most recent GPU ConvNet implementations with standard frameworks.
The measurements on Tesla P100 have been performed in \cite{NVIDIA-WP17}. Geforce GTX780 and NVIDIA Tegra K1 have been directly used and measured in our group \cite{LUKAS-DAC15}. Tegra X1 has been used in \cite{NVIDIA-WP15}.
Measurements on Tesla K20, K40, Geforce GTX Titan, and GTX 770 have been performed by Berkely AI Lab \cite{CAFFE-GPU}. Finally, Geforce GTX480 and  NVIDIA GT335m has been used in \cite{NEUFLOW-ASIC}.
A couple of points are interesting to observe in this plot: we saw previously on page \pageref{fig:ht7-8} that SMC achieves an almost uniform performance for all studied ConvNets (more than 90\% efficiency) while looking at the results of Tesla P100 we can observe more variations. This is mainly thanks to the hardware implementation of nested loops in the NeuroStream computation paradigm, and to the availability of a large amount of DRAM at a low latency in SMC.
Furthermore, the SMC achieves about 3.5X improvement when compared with GPUs in similar technologies (e.g. GTX780, Tegra K1, Tesla K40). Compared to Tegra X1 with 20nm technology it achieves 2.7X, and compared to Tesla P100 (16nm) for GoogLeNet and ResNet, on the average, it achieves a similar energy efficiency. For AlexNet, however, Tesla P100 achieves 1.5X better efficiency compared to our solution.
Another glance at the above plot reveals that ConvNets cannot easily exploit the maximum throughput offered by the GPUs. We can observe that even on the most powerful GPUs, the utilization of computing resources does not exceed 55\%, which is directly reflected in a lower energy-efficiency.
In fact, they are optimized to perform general matrix multiplication (GEMM) operations, and ConvNets should be transformed into these forms for execution on the GPU platforms \cite{LUKAS-DAC15}. However, for modern ConvNets with growing memory footprints and non-uniform layers it is becoming more difficult (and wasteful in terms of memory footprint and bandwidth) to transform them into GEMM formats. This is in contrast with our proposal which executes ConvNets with more than 90\% performance efficiency for all studied ConvNets (See \autoref{perf}).
Also, the aggregate bandwidth available in the multi-SMC scenario is much higher than what can be delivered to the host processors and accelerators. This makes our solution more scalable in comparison with high-performance FPGAs and GPUs.

\section{Conclusions and Ongoing Work} \label{con}

In this paper, we proposed a scalable and energy-efficient PIM system based on a network of multiple SMC devices. 
Each SMC is augmented with a flexible clustered many-core called NeuroCluster, capable of executing deep ConvNets with growing memory footprints and computation requirements.
To this end, NeuroStream (a streaming FP32 coprocessor) is proposed, along with an efficient tiling mechanism and a scalable computation paradigm.
Our proposal increases the LoB die area of the standard HMC only by 8\%, and achieves an average performance of 240\,GFLOPS for complete execution of full-featured modern ConvNets within a power-budget of 2.5\,W.
22.5\,GFLOPS/W energy efficiency is achieved in a single SMC (consuming 11\,W in total) which is 3.5X better than the best GPU implementations in similar technologies. The performance was shown scalable with a network of SMCs.
It is worth stressing that our platform allows for offloading the ConvNet tasks completely into the memory cubes at a minor system power cost. This implies that the compute logic in the host SoC is free to deal with other workloads. Also, the cost increase with respect to a baseline HMC system would be negligible. Therefore, essential ConvNet acceleration is provided at a very small system cost.
Ongoing research efforts include silicon implementation of the NeuroCluster with 5 clusters, parallel implementation of training on this architecture, and pushing further to achieve higher performance and efficiency inside the cubes (e.g. more advanced refresh and power management schemes to reduce the power in unused DRAM pages).

\ifCLASSOPTIONcaptionsoff
\newpage
\fi

\bibliographystyle{IEEEtran}
\bibliography{erfan-paper}{}

\begin{IEEEbiography}[{\includegraphics[height=1in,clip,keepaspectratio]{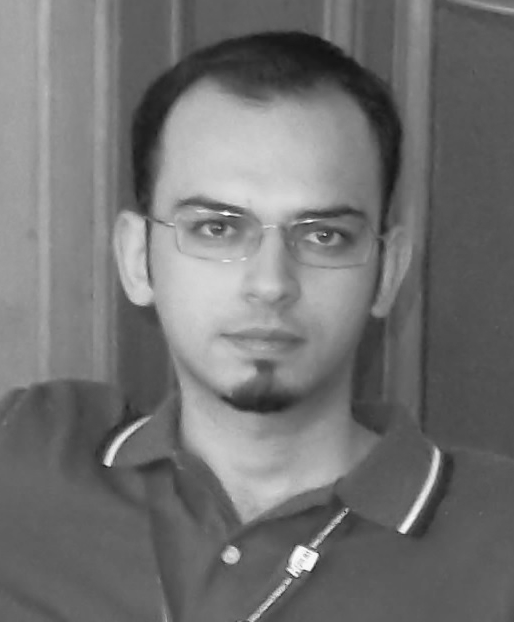}}]{Erfan Azarkhish}
	received the B.Sc and M.Sc. degrees from the %
	University of Tehran, Tehran, Iran in 2007 and 2009, respectively. He received his Ph.D. degree in Electronics Engineering from the University of Bologna, Bologna, Italy, in 2015, and served as a post-doc researcher there until 2017. He currently holds a senior R\&D engineer position at the Swiss Center for Electronics and Microtechnology, Neuch\^{a}tel, Switzerland.
	His main research interests are neuromorphic computing and near-memory processing.
\end{IEEEbiography}

\begin{IEEEbiography}[{\includegraphics[height=1in,clip,keepaspectratio]{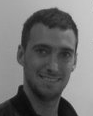}}]{Davide Rossi}
received the PhD from the University of Bologna, Italy, in 2012. He has been a postdoc researcher in the Department of Electrical, Electronic and Information Engineering at the University of Bologna since 2015, where he currently holds an assistant professor position. His research interests focus on energy efficient digital architectures in the domain of heterogeneous and reconfigurable multi and many-core systems. In this fields he has published more than 50 paper in international peer-reviewed conferences and journals.
\end{IEEEbiography}

\begin{IEEEbiography}[{\includegraphics[height=1in,clip,keepaspectratio]{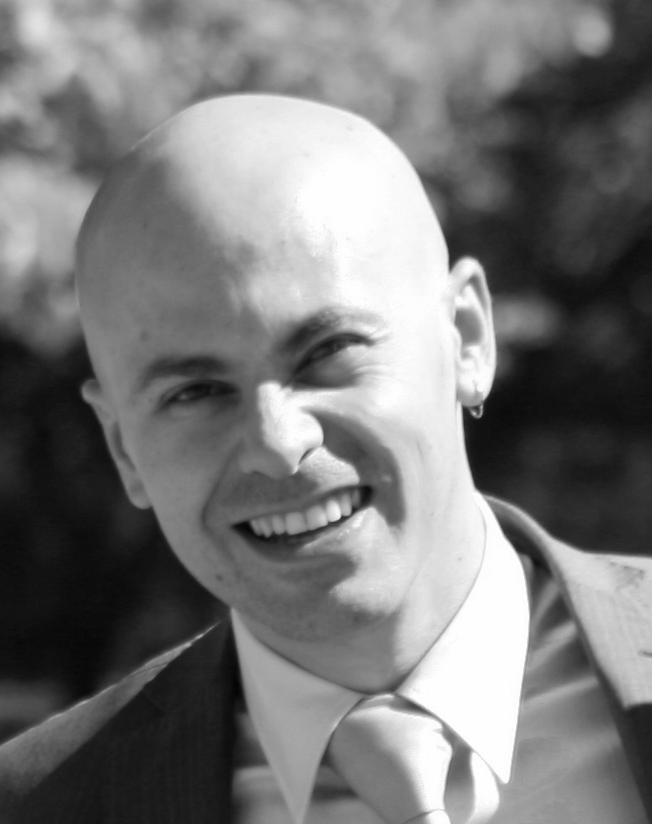}}]{Igor Loi}
	received the B.Sc. degree in Electrical Engineering from the University of Cagliari, Cagliari, Italy, in 2005 and the Ph.D. degree in the Department of Electronics and Computer Science, University of Bologna, Italy, in 2010. He is currently holding the position of Assistant Professor in Electronic Engineering at the University of Bologna. He is focusing his research on ultra-low power multi-core systems, memory systems evolution, and ultra low-latency interconnects.
\end{IEEEbiography}

\begin{IEEEbiography}[{\includegraphics[height=1in,clip,keepaspectratio]{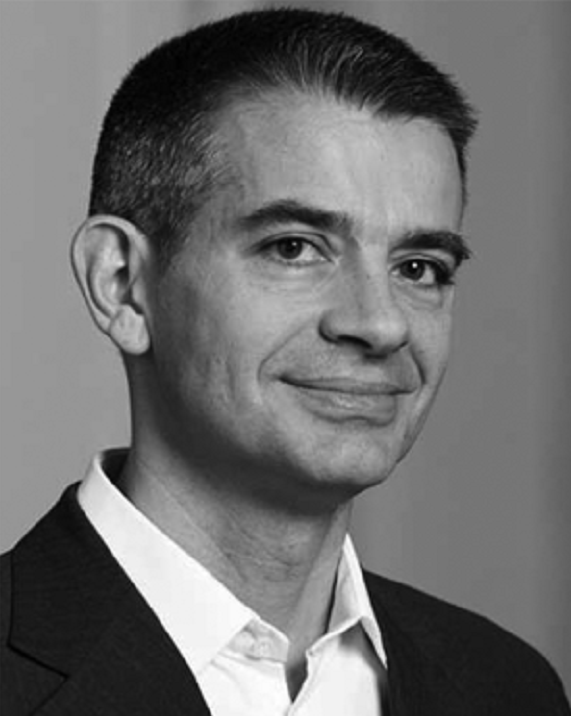}}]{Luca Benini}
	is the Chair of Digital Circuits and Systems at ETH Zurich and a Full Professor at the University of Bologna. He has served as Chief Architect for the Platform2012/STHORM project at STmicroelectronics, Grenoble. Dr. Benini's research interests are in energy-efficient system and multi-core SoC design. He has published more than 800 papers in peer-reviewed international journals and conferences, four books and several book chapters. He is a Fellow of the ACM and a member of the Academia Europaea.
\end{IEEEbiography}

\end{document}